\documentclass[letterpaper, conference, dvipsnames]{IEEEtran}

\usepackage{subfiles}
\usepackage[english]{babel}
\usepackage{graphicx}

\usepackage[utf8]{inputenc}
\usepackage[colorlinks, unicode=true, backref=section]{hyperref}

\usepackage{pgffor}
\usepackage{wrapfig}
\usepackage{enumerate}
\usepackage[square,sort,comma,numbers]{natbib}
\usepackage{url}
\usepackage{mathtools}
\usepackage{afterpage}
\usepackage{bm}
\usepackage{tabu}
\usepackage{adjustbox}
\usepackage{amsthm}
\usepackage{enumitem}
\usepackage{tikz}
\usepackage{pgfplots}
\usepackage{pgffor}
\usepackage{calculator}
\usepackage{cleveref}
\usepackage{adjustbox}
\usepackage{amsmath, amssymb, amsfonts}
\usepackage{contour}
\usepackage{paralist}
\usepackage{soul}
\usepackage{authblk}

\usepackage{textcomp} 
\usepackage{stmaryrd}

\setdefaultleftmargin{\parindent}{}{}{}{}{}

\usetikzlibrary {positioning,graphs,calc,decorations.pathmorphing,shapes,arrows.meta,arrows,shapes.misc}
\usetikzlibrary{fit}
\usetikzlibrary{shapes.symbols,decorations.text}
\usetikzlibrary{fadings}
\usetikzlibrary{decorations.pathmorphing}
\usetikzlibrary{decorations.markings}
\usetikzlibrary{decorations.pathreplacing}
\usetikzlibrary{calc}
\usetikzlibrary{backgrounds,patterns}
\usetikzlibrary{intersections}

\definecolor{nodecolor}{rgb}{0.03, 0.91, 0.87}
\definecolor{level1color}{rgb}{0.6, 0.98, 0.6}
\definecolor{level2color}{rgb}{0.31, 0.78, 0.47}
\definecolor{level3color}{rgb}{1.0, 0.87, 0.0}
\definecolor{level4color}{rgb}{1.0, 0.55, 0.41}
\def\grisclair{gray!40}
\def\grisfonce{gray!100}

\definecolor{yaleblue}{rgb}{0.06, 0.3, 0.57}
\definecolor{vermilion}{rgb}{0.89, 0.26, 0.2}

\pgfplotscreateplotcyclelist{hancol}{
{yaleblue},
{vermilion},
}

\theoremstyle{definition}

\newtheorem{thm}{Theorem}
\newtheorem{thm_bis}{Theorem}

\theoremstyle{remark}
\newtheorem{rmk}{Remark}

\newcommand{\tb}[1]{\textbf{#1}}
\newcommand{\ti}[1]{\textit{#1}}
\newcommand{\red}[1]{{\color{red}$\blacksquare$ #1}}

\newcommand{\sanFermin}{San Ferm\'{i}n}

\newcommand{\inc}{\mathit{In}} 
\newcommand{\out}{\mathit{Out}} 
\newcommand{\unv}{\mathit{Uc}} 

\newcommand{\byz}{b}
\newcommand\vpl{VP}
\newcommand{\Part}[2]{P^{#1}_{#2}}   

\newcommand{\calP}{\mathcal{P}}
\newcommand{\calB}{\mathcal{B}}

\newcommand{\outcomm}{\lceil Cn\rceil}

\newcommand{\proba}{\mathbb{P}}

\newcommand{\start}[1]{StartTime_{#1}}
\newcommand{\vp}{\mathrm{VP}}

\title{Handel: Practical Multi-Signature Aggregation for Large Byzantine Committees}

\author[1]{Olivier B\'egassat\thanks{olivier.begassat@consensys.net}}
\author[2]{Nicolas Gailly\thanks{nicolas.gailly@protocol.ai}}
\author[1]{Blazej Kolad\thanks{blazej.kolad@consensys.net}}
\author[1]{Nicolas Liochon\thanks{nicolas.liochon@consensys.net}}
\affil[1]{ConsenSys R\&D}
\affil[2]{Protocol Labs}

\date{}

\pgfplotsset{
    compat=1.14,
    tick align = outside,
    scaled y ticks=false, 
    scaled x ticks=false, 
    yticklabel style={/pgf/number format/fixed}, 
    xticklabel style={/pgf/number format/fixed}, 
    label style={font=\scriptsize},
    tick label style={font=\scriptsize},
    legend style={font=\tiny},
    discard if not/.style 2 args={
        x filter/.code={
            \edef\tempa{\thisrow{#1}}
            \edef\tempb{#2}
            \ifx\tempa\tempb
            \else
                \def\pgfmathresult{inf}
            \fi
        }
    }
}

\tikzstyle{dashed1}=          [mark=]
\tikzstyle{dashed2}=          [thick, dash pattern=on 12pt off 1pt,mark=]
\tikzstyle{dashed3}=          [thick, dash pattern=on 6pt off 2pt,mark=]
\tikzstyle{dashed4}=          [thick, dash pattern=on 1pt off 1pt,mark=]
\tikzstyle{dashed5}=          [thick, dash pattern=on 4pt off 9pt,mark=]
\tikzstyle{dashed6}=          [thick, dash pattern=on 1pt off 9pt,mark=]

\begin{document}
\maketitle
\thispagestyle{plain}
\pagestyle{plain}

\begin{abstract}
We present Handel, a Byzantine-tolerant aggregation protocol that allows for the quick aggregation of cryptographic signatures over a WAN. Handel has polylogarithmic time, communication and processing complexity. We implemented Handel as an open source Go library with a flexible design to support any associative and commutative aggregation function. We tested Handel on 2000 AWS instances running two nodes per instance and located in 10 AWS regions. The 4000 signatures are aggregated in less than 900 milliseconds with an average per-node communication cost of 56KB.
\end{abstract}

\section{Introduction}\label{sec:intro}
Many large scale decentralized systems need to efficiently establish agreement between thousands of untrusted nodes. For instance, in the blockchain protocols Algorand~\cite{gilad2017algorand}, Dfinity~\cite{hanke2018dfinity}, Tendermint~\cite{buchman2016tendermint} and Ethereum 2~\cite{eth2}, nodes agree on a common message by exchanging and collecting votes (cryptographic signatures) expressing the agreement of individual nodes.

A common practical solution which bypasses large scale vote collection is to delegate the vote to committees, e.g. small subsets of randomly selected nodes.
Thus, in Dfinity committees generate threshold signatures for block notarization. However, there are issues with using committees in a Byzantine context. One issue is that of committee selection. In the adaptive adversaries model~\cite{canetti2001adaptive}, committee selection must be private and unpredictable. If the randomness is manipulable, Byzantine nodes may take control of future committees. Moreover, if committee members and their IP addresses are public, as they are in Dfinity, they become high profile targets for attackers. Lastly, committees are far more vulnerable than the network as a whole. Indeed, the resources needed to compromise a committee are, by construction, orders of magnitude lower than those needed to compromise the network as a whole.
Bribing attacks~\cite{Bonneau2016WhyBW}, for instance, become a possibility. Also, having private and unpredictable committees is no protection against bribing, as demonstrated by public and trustless bribing contracts~\cite{McCorry2018SmartCF}.
Hence, we argue that supporting large committees, or eschewing committees altogether, allows existing public blockchain networks to significantly increase their security and/or to scale in size while maintaining security.

Multi-signature schemes, first formally described by Micali et al. in~\cite{Micali2001}, allow a group of users to produce aggregate signatures that are both smaller and faster to verify than the corresponding collection of individual signatures. 
The BLS multi-signature scheme~\cite{boldyreva2002}, derived from Boneh, Lynn, Schacham's 2001 elliptic curve cryptography signature scheme~\cite{boneh2001short}, achieves \textit{constant size multi-signatures} and \textit{constant time verification} and does not require a setup. Its aggregation function is both \textit{commutative} and \textit{associative}.

When using such a multi-signature scheme, the problem of vote collection reduces to one of aggregation at scale.
Large scale aggregation protocols have been studied \cite{gupta2001scalable,kempe2003gossip,yalagandula2004sdims,li2005implementing}, but there are, to date, no scalable, Byzantine-tolerant~\footnote{%
Since "Practical Byzantine Fault Tolerance" by Castro and Liskov~\cite{castro1999}, the term "Byzantine fault tolerance" is often associated with consensus. Machine Learning practitioners use the term Byzantine-tolerance~\cite{blanchard2017}. We adopted the same wording.} aggregation protocols working over a WAN. 
Cappos et al.'s aggregation protocol \sanFermin{}~\cite{cappos2008san} achieves time and communication complexity that is logarithmic relative to node count when there is no failure, but it is not Byzantine-tolerant. Furthermore, its reliance on timeouts means that with short timeouts and high network latency, valid contributions will be ignored, but long timeouts slow the protocol down. \sanFermin{} thus works well in permissioned environments with homogeneous networks, but is inadequate for fully decentralized systems. 

\subsection{High level presentation of the protocol}
Handel is an aggregation protocol that provides \textbf{Byzantine-tolerance} and \textbf{versatility} (it handles heterogeneous network latencies and computer capacities), along with \textbf{speed} (it aggregates thousands of contributions in seconds) and \textbf{efficiency}.
Handel assumes that the whole list of participants is known when starting the protocol. Nodes organize themselves as the leaves of a binary tree and aggregate in parallel. To avoid remaining stuck on low levels due to slow or unresponsive nodes, each node aggregates on all levels concurrently, optimistically sending its current levelwise best aggregate contributions, even if incomplete. However, they do not contact all their peers simultaneously, but rather one by one, periodically. In a Byzantine context, nodes may receive invalid contributions; contributions are therefore verified before they are aggregated. As verification is costly, nodes \textit{score} contributions and \textit{prune} redundant ones.
Finally, Handel uses peer ranking plus a scoring and windowing mechanism as a defense against DoS attacks.

\subsection{Practical use case}\label{sec:intro:depl}
Handel's main use case is as a drop-in component for committee-free vote aggregation in proof-of-stake blockchains. In Ethereum~2 (\cite{eth2}, currently under implementation), the  complete list of validators is known, committees are selected by means of a shared source of randomness and blocks are generated at fixed intervals (one every 6 seconds); the signature scheme is BLS. Some critical state changes require multiple votes from consecutive committees, leading to a time to finality of about 13 minutes (two epochs in Ethereum~2 vocabulary).
Using Handel for attestation aggregation would allow the entire validator set to participate in a single vote.
The Ethereum~1 network currently encompasses 8,500 nodes\footnote{cf. \href{https://www.ethernodes.org/network/1}{https://www.ethernodes.org/network/1}}; proponents of Ethereum~2 plan for the network to eventually gather tens of thousands of validator nodes (Ethereum~2 ultimately targets 400,000 \textit{logical} validators, not all of them being physical nodes).
Our simulations show that with 32,000 nodes, Handel's aggregation time is far shorter than the 6 second 
block-time: we measured an actual aggregation time of 1.2\textit{s} with 25\% of nodes being fail-silent (cf. figure~\ref{cpx-silent}).
For Ethereum~2 finality, which requires two votes, this would bring time to finality down to 18 seconds. A reduction of the block time to 3 seconds is conceivable and would lead to a time to finality of 9 seconds.

Incidentally, Handel can be used as an aggregation protocol in contexts that do not require Byzantine-tolerance, but benefit from its versatility.
Systems implemented in garbage-collected languages, such as Java or Go, are prone to latency spikes, making timeout configuration difficult. Carpen-Amarie et al.~\cite{carpen2015performance} found that "even the concurrent low-pause [garbage collectors] end up stopping the application threads for a few seconds".
HBase~\cite{Dimiduk2012HBaseIA}, a Java implementation of the Bigtable~\cite{chang2008bigtable} distributed database whose clusters can contain thousands of nodes, is an example of such a system.
HBase supports server-side code through its ''co-processor'' mechanism and the nodes' list is accessible to the client applications, making such an integration possible.

\subsection{Implementation}
We implemented Handel as an open-source Go library~\cite{handellib}.
Our implementation allows users to plug different aggregation schemes (multi-signature or otherwise). We implemented extensions to use Handel with BLS multi-signatures. Security, signature size and signature verification time depend on the underlying curve. We used the BN256 curve~\cite{bn256cf,barreto2005pairing}: (compressed) signatures are 256 bits, verification takes a few milliseconds~\cite{mikuli}. We ran large-scale tests and evaluated Handel on 2000 AWS nano instances, each running two Handel nodes and located in 10 AWS regions. In one series of tests with a threshold of 99\% and 4000 signatures to aggregate, Handel terminated on average in under 900 milliseconds, with an average per-node communication cost of 56KB.

\subsection{Our Contributions}
We contribute the Handel protocol (section~\ref{sec:protocol}) and an open source implementation (section~\ref{sec:impl}). Our analysis (section~\ref{sec:impl}) includes a detailed proof of its convergence (appendix ~\ref{app: proof of convergence}). We also study  the different anonymity options (section~\ref{sec:anonymity}). Finally we evaluate Handel in a large scale context  (section~\ref{sec:eval}). 

\section{Model}\label{sec:model}
\subsection{Contributions}\label{sec:model:contributions}
Handel's purpose is to aggregate pieces of data we call \textbf{contributions}. The exact nature of the contribution and the aggregation function is defined by the application.
For instance, contributions could be numbers, and aggregation summing.
More formally, consider a nonempty set $C$ whose elements are called \textbf{contributions}, a set $\mathit{Pub}$ of \textbf{public data},
a \textbf{verification function} $\mathit{Ver}^\mathit{Pub}:C\to\{0,1\}$ that verifies contributions against public data, a \textbf{(partial) aggregation function}
$\mathit{Aggr}:C\times C\nrightarrow C$ that performs aggregation of contributions and a \textbf{weight function} $w:C\to\mathbb{N}$ to compare contributions.
A contribution $c\in C$ is \textbf{valid} against the public data $\mathit{Pub}$ if and only if $\mathit{Ver}^\mathit{Pub}(c)=1$. In practice, not all pairs of contributions $(c,c')$ may be meaningfully aggregated, hence we insist aggregation be a \textit{partial} function (i.e. defined on a subset of $C\times C$).
Define a pair $(c,c')$ of contributions to be \textbf{aggregable} if
$(c,c')$ is in the domain of the partial function $\mathit{Aggr}$, and an \textbf{aggregate contribution}
to be a contribution in the image of $\mathit{Aggr}$. Contributions that are not aggregate contributions are called \textbf{individual contributions}. Aggregate valid contributions include any individual contribution at most once. 
We require that the (partial) aggregation function satisfies both \textit{commutativity} and \textit{associativity} conditions. The weight function measures how many individual contributions are contained in a given contribution. Given two contributions, we call \textbf{best} contribution the weightier one.
We consider that contribution verification is the only CPU intensive operation.

An \textbf{aggregation protocol} is a protocol for a set of connected participants (an \textbf{aggregator set}), each owning an individual contribution, to aggregate these until either a threshold is reached for all participants or a time limit is reached.

\subsection{Participants}\label{sec:model, participants}
There are $N$ participants in the system. Each participant has a unique public key and network address. During its setup phase, Handel creates for each participant $P$ a unique ID $\mathit{id}_P \in\llbracket0,N\llbracket\,$; we henceforth conflate participants with their unique identifier.

Participants are connected by point-to-point \ti{authenticated} channels. The \tb{network} is \ti{partially synchronous} as defined in~\cite{dwork1988consensus}: there is a fixed (but unknown) bound on network latency. Message delivery is guaranteed.

Each participant: (1) knows all other participants as well as their public keys and network address, (2) knows its own contribution when the protocol starts, (3) has an internal clock, (4) has access to a public source of randomness. We assume a \ti{setup phase} where this shared knowledge is established. Byzantine nodes participate in the setup phase, but cannot prevent the global knowledge from being established. We assume that all participants start the protocol at roughly the same time, i.e. the maximum delay
between two honest participants' start time is less than the maximum network latency\footnote{In practice, it means that participants should run NTP: Murta and al. measured in 2005 a mean delta of 7 milliseconds~\cite{MurtaTM06}, far less than the typical latency on a WAN. NTP is a quite common requirement. For example, Ethereum 1 or HBase already recommend using NTP.}.

The application starts the protocol either (1) at predetermined times (e.g. Ethereum~2) or (2) when a trusted server triggers the application (e.g. HBase). The application runs the protocol either (1) for a predetermined duration (e.g. 3 seconds, as in Ethereum~2) or (2) until at least one honest node reaches the threshold and communicates it (e.g. a new block was created by the blockchain).

\subsection{Threat model}\label{sec:model:threat}
The adversary controls a \ti{static subset} of the participants. These Byzantine participants behave \textit{arbitrarily} during the execution of the protocol and \ti{coordinate among themselves}.

The adversary can read the full state of any participant, and the data exchanged over any communication channel. However the adversary cannot directly modify the state of the honest participants, impersonate them, or interfere in the communications between honest participants. The adversary's messages are not delayed and arrive instantaneously.

Participants are bound to \ti{polynomial-time computation}, e.g. cannot break commonly used cryptographic primitives. Specifically they cannot forge valid contributions. Byzantine participants may send false contributions that will fail verification, wasting receiver's resources. Thanks to their omniscience, Byzantine participants may create any \ti{valid} aggregate contribution they wish.

\subsection{Goals}
Handel outputs an aggregate contribution including at least a predefined threshold number $T$ of contributions, $T$ being no larger than the number of honest nodes. Handel guarantees the following:
\begin{compactitem}
    \item \tb{Safety:} Handel never generates an invalid aggregate contribution.
    \item \tb{Termination:}
    Handel eventually finishes with an aggregate contribution including at least a configuration-defined threshold number of individual contributions.
    \item \tb{Time efficiency:} Handel's completion time scales poly-logarithmically with the number of participants.
    \item \tb{Resource efficiency:}  Handel's CPU and bandwidth consumption scale polylogarithmically with the number of participants.
\end{compactitem}
Handel does not guarantee uniformity of results: participants can reach the threshold with different sets of contributions. In other words, Handel is \textit{not} a consensus protocol.

\subsection{Application to BLS}
For concreteness, let us describe the components of contributions in the case of BLS signature aggregation~\cite{boneh2001short,boldyreva2002}. The $N$ participants create BLS signatures for a message $m$. We define $C$ to be $E\times(\{0,1\}^N\setminus\{[0,\dots,0]\})$ where $E$ is the subgroup of the elliptic curve used for BLS signatures~\cite{boneh2001short}. Thus, a contribution is a pair $(\sigma,[\epsilon_0,\dots,\epsilon_{N-1}])$ where $\sigma\in E$ is a point on the curve (an aggregate signature), and $[\epsilon_0,\dots,\epsilon_{N-1}]$ is an ordered list of bits (not all zero).
This ordered list of bits is used to keep track of whose signatures are (supposedly) included in $\sigma$: participant $i$'s signature (i.e. associated with $pk_i$) ought to be ``accounted for in $\sigma$'' iff $\epsilon_i=1$.
The public data is the set
$\mathit{Pub}=\{m,[pk_0,\dots,pk_{N-1}]\}$
containing the message $m$ and the
ordered list of the participants' public keys.
The verification function is defined
on $c=(\sigma,[\epsilon_0,\dots,\epsilon_{N-1}])$ by
\[
	\mathit{Ver}^\mathit{Pub}
	(c)
	=
	\mathrm{IsOne}
	\left(
	e(\sigma,g)^{-1}
	\cdot
    e
    \Big(
    H(m),
    \prod_{i=0}^{N-1}
    pk_i^{\epsilon_i}
    \Big)
	\right)
\]
where $e:E\times E\to\mathbb{F}^\times$ is the pairing used in the BLS scheme (with values in the multiplicative group $\mathbb{F}^\times$ of some finite field), $g\in E$ is the chosen generator used to construct public keys from secret keys, $H$ is a hash function from the message space onto $E$ and $\mathrm{IsOne}:\mathbb{F}^\times\to\{0,1\}$ is constant equal to $0$ except for $\mathrm{IsOne}(1)=1$. Thus, a contribution is valid if and only if it is the product of the 
signatures on $m$ 
by the public keys represented in the bitset $[\epsilon_0,\dots,\epsilon_{N-1}]$.

The domain of the (partial) aggregation function $\mathit{Aggr}$ for BLS signatures consists of all pairs of contributions
$c=(\sigma, [\epsilon_0,\dots,\epsilon_{N-1}])$ and 
$c'=(\sigma', [\epsilon_0',\dots,\epsilon_{N-1}'])$ whose bitsets are ``disjoint'' (i.e. for all $i$,
$\epsilon_i\epsilon_i'=0$). We set
\[
    \mathit{Aggr}(c,c')=(\sigma+_E\sigma',[\epsilon_0+\epsilon_0',\dots,\epsilon_{N-1}+\epsilon_{N-1}'])
\]
(where $+_E$ is point addition on the elliptic curve). Aggregate contributions are precisely those contributions whose bitset contains at least two nonzero bits, while individual contributions are those contributions whose bitset contains precisely one nonzero bit. 
We define the weight of a contribution $c=(\sigma,[\epsilon_0,\dots,\epsilon_{N-1}])$ to be $w(c)=\epsilon_0+\cdots+\epsilon_{N-1}$.

\subsection{Simulation}\label{sec:model:simulation}
We simulated Handel to determine the best parameter values and to test Handel's resilience against various Byzantine attacks. Our model for these simulations (1) uses real world latencies as measured by the WonderNetwork project \cite{latencyInternet} (WonderNetwork provides the mean latencies for 242 cities) (2) nodes are distributed randomly across these 242 cities weighted by population density (3) nodes start the protocol with uniformly distributed delay $\in[0\mathit{ms},100\mathit{ms}]$  (4) verification time is 4ms on a standard node, but nodes can be up to 3 times slower or up to 3 times faster, with verification times following a truncated Gaussian distribution.
We measured the average time for the honest nodes to reach a threshold of 99.9\% of the honest nodes. Our results are averages of 5 runs. The code of the simulator is open source~\cite{wittgenstein}.
\section{Handel protocol}\label{sec:protocol}
This section introduces the techniques used by Handel, then details how Handel works.
Some parameters are given explicit values; we explain those choices in section~\ref{sec:rationale}.

\subsection{Overview of the techniques}\label{sec:handel:presentation}
Handel combines multiple techniques to reach its goals. A standard technique for aggregation has the nodes organize themselves as the nodes of a binary tree. However, this is not fault tolerant: a node failure leads to the loss of the contributions from all the nodes in the subtree under it.

Thus, Handel uses San Ferm\'{i}n's \textit{tree based overlay network}: nodes organize themselves by node ID as the \textit{leaves} of a binary tree and can connect to all other nodes.
Aggregation is thus parallelized, providing the basis for logarithmic time complexity.
Connections are structured in levels.
Assume for simplicity that $N$ is a power of $2$.
At the first level, every node $n$ has a single peer $n'$, and $n'$ has $n$ as its level 1 peer.
At the second level, each such pair of nodes $n$, $n'$ has two peers 
$m$, $m'$, themselves being level 1 peers. Again, this is symmetrical.
The four nodes $n$, $n'$, $m$, $m'$ have four level 3 peers and so forth.

With IDs being used to assign positions in the overlay network, Byzantine nodes could try to choose their position in the tree by carefully choosing their ID. To prevent position-based attacks, IDs are shuffled at the beginning of the protocol.

To avoid remaining stuck on low levels due to slow or unresponsive nodes, Handel nodes aggregate on all levels concurrently: nodes optimistically send their current levelwise best aggregate contributions, even if incomplete.
This brings fail-silent fault tolerance but threatens to flood the network.

To circumvent this, nodes engage in \textit{periodic dissemination}. 
They do not contact all their peers simultaneously, but rather one by one,
periodically and in parallel on all active levels. To save on communication complexity, not all levels are active when the protocol starts: nodes activate a new level (the next level in ascending order) every 50 milliseconds.

In a Byzantine context, nodes may receive invalid contributions; contributions should therefore be verified before they are aggregated.
But verifying contributions can be costly (a few milliseconds for BLS signatures~\cite{mikuli}): nodes with limited CPU capacity may find themselves unable to verify all incoming contributions immediately and having to buffer them.
Furthermore, nodes may receive multiple contributions at a given level, some of which may prove redundant, inferior to the node's current best contribution for that level or less valuable than others. Therefore, nodes \textit{score} contributions and \textit{prune} redundant ones before verifying them.
This brings \textit{resource efficiency}---redundant contributions are not verified---and \textit{versatility}---slow nodes need not verify all contributions.

However, this also creates an attack point:
in an effort to waste other nodes' CPU resources,
Byzantine nodes could flood the network with high scoring, yet invalid contributions.
In the worst case, honest nodes may find themselves wasting most of their resources verifying invalid contributions.
Every Handel node thus prioritizes incoming contributions by means of a \textit{local ranking} of its peers, independent of received contributions.
The ranking is used to determine the subset (the \textit{window}) of peers in a level whose contributions will be scored. \ti{Window size} is dynamic: when a verification fails, window size decreases; when a verification succeeds, window size increases.
When several verifications succeed in a row,
the window gets wider and Handel can freely select
which contribution to verify.
When several verifications fail in a row, the window may become very narrow; in this case, the set of contributions Handel considers for verification is small. However, an attacker can no longer force its contributions to be verified.
To allow a node to aggregate individual contributions even if the aggregate contribution is not \ti{aggregable} with its local aggregate contribution, messages must include the sender's \textit{individual contribution} alongside the relevant aggregate one.


When a node has aggregated all contributions it can receive for a level, it need not verify any further contributions from that level. To leverage this, when a node manages to assemble a complete aggregate contribution of a level, it \textit{immediately} sends it to multiple nodes in the relevant peer set without waiting for periodic dissemination. This mechanism is called \ti{fast path}.

All communication (i.e. periodic dissemination and fast path) is
\textit{one-way}: nodes \textit{push} messages. To avoid complex questions around distinguishing unresponsiveness from Byzantine behaviour, Handel eschews any timeout-based mechanism.

\subsection{Tree based overlay network}\label{sec:handel:tree}
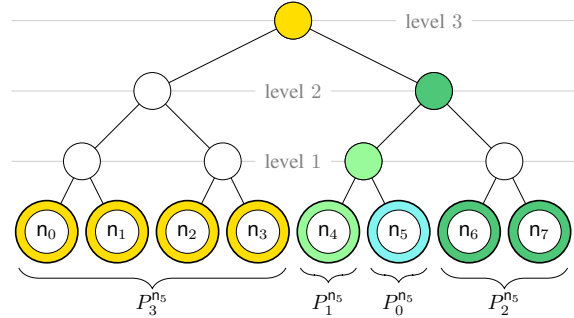
\begin{figure}[!ht]
    \centering
    \scalebox{0.78}{\trimbox{1.2cm 0cm 0cm 0cm}{\pgfmathsetmacro\vert{1.2}
\pgfmathsetmacro\hori{1.2}
\pgfmathsetmacro\outer{.53}
\pgfmathsetmacro\inner{21}
\pgfmathsetmacro\factor{.90}

\begin{tikzpicture}
\def\n{3}

\draw[color = \grisclair]
        (0,0)
            -- node [circle, fill=white, pos=.75] {\color{\grisfonce}level $3$}
                (8*\hori,0);
\draw[color = \grisclair]
        (0,-\vert)
            -- node [circle, fill=white] {\color{\grisfonce}level $2$}
                (8*\hori,-\vert);
\draw[color = \grisclair]
        (0,-2*\vert)
            -- node [circle, fill=white] {\color{\grisfonce}level $1$}
                (8*\hori,-2*\vert);
        
\foreach \i in {0,...,\n} {
    \pgfmathsetmacro\TwoToTheI{2^\i}
    \pgfmathsetmacro\TwoToTheNMinusI{2^\n/2^\i}
    \foreach \j in {1,...,\TwoToTheI} {
        \path
            (\TwoToTheNMinusI*\j*\hori-.5*\TwoToTheNMinusI*\hori,-\i*\vert)
                coordinate
                    (vertex=\i=\j);
    }
}
\pgfmathsetmacro\nRemoveOne{int(\n-1)}
\foreach \i in {0,...,\nRemoveOne} {
    \pgfmathsetmacro\TwoToTheI{2^\i}
    \foreach \j in {1,...,\TwoToTheI} {
        \pgfmathsetmacro\nexti{int(\i+1)}
        \pgfmathsetmacro\l{int(2*\j-1)}
        \pgfmathsetmacro\r{int(2*\j)}
        \pgfmathsetmacro\Inti{int(\i)}
        \pgfmathsetmacro\Intj{int(\j)}
        \draw (vertex=\Inti=\Intj) -- (vertex=\nexti=\l);
        \draw (vertex=\Inti=\Intj) -- (vertex=\nexti=\r);
    }
}
\draw[draw=black, thick, fill=nodecolor!50] (vertex=\n=6) circle(\outer);
\draw[draw=black, thick, fill=level1color] (vertex=\n=5) circle(\outer);
\draw[draw=black, thick, fill=level2color] (vertex=\n=7) circle(\outer);
\draw[draw=black, thick, fill=level2color] (vertex=\n=8) circle(\outer);
\draw[draw=black, thick, fill=level3color] (vertex=\n=1) circle(\outer);
\draw[draw=black, thick, fill=level3color] (vertex=\n=2) circle(\outer);
\draw[draw=black, thick, fill=level3color] (vertex=\n=3) circle(\outer);
\draw[draw=black, thick, fill=level3color] (vertex=\n=4) circle(\outer);

\foreach \i in {0,...,\nRemoveOne} {
    \pgfmathsetmacro\TwoToTheI{2^\i}
    \pgfmathsetmacro\TwoToTheNMinusI{2^\n/2^\i}
    \foreach \j in {1,...,\TwoToTheI} {
        \pgfmathsetmacro\number{int(\TwoToTheI+\j-1)}
        \pgfmathsetmacro\Inti{int(\i)}
        \pgfmathsetmacro\Intj{int(\j)}
        \draw (vertex=\i=\j) node [circle, draw = black, fill = white] (vertex=\Inti=\Intj)  {\phantom{5}};
    }
}
\pgfmathsetmacro\TwoToTheN{2^\n}
\foreach \j in {1,...,\TwoToTheN} {
    \pgfmathsetmacro\Intj{int(\j)}
    \pgfmathsetmacro\number{int(\j-1)}
    \draw (vertex=\n=\j) node [circle, draw = black, fill = white, inner sep=0pt, minimum size = \inner] {$\textsf{n}_{\number}$};
}
\path   (vertex=\n=6) ++ (-\factor*\outer, -\factor*\outer)
            coordinate (level -1 left) ;
\path   (vertex=\n=6) ++ (\factor*\outer, -\factor*\outer)
            coordinate (level -1 right) ;
\draw[decorate,decoration={brace,amplitude=7pt,mirror,raise=4pt},yshift=0pt]
    (level -1 left) -- (level -1 right)
        node[black,midway, yshift = -.75cm]
            {$\displaystyle \Part{\textsf{n}_5}{0}$};
\path   (vertex=\n=5) ++ (-\factor*\outer, -\factor*\outer)
            coordinate (level 0 left) ;
\path   (vertex=\n=5) ++ (\factor*\outer, -\factor*\outer)
            coordinate (level 0 right) ;
\draw[decorate,decoration={brace,amplitude=7pt,mirror,raise=4pt},yshift=0pt]
    (level 0 left) -- (level 0 right)
        node[black,midway, yshift = -.75cm]
            {$\displaystyle \Part{\textsf{n}_5}{1}$};
\path   (vertex=\n=7) ++ (-\factor*\outer, -\factor*\outer)
            coordinate (level 1 left) ;
\path   (vertex=\n=8) ++ (\factor*\outer, -\factor*\outer)
            coordinate (level 1 right) ;
\draw[decorate,decoration={brace,amplitude=10pt,mirror,raise=4pt},yshift=0pt]
    (level 1 left) -- (level 1 right)
        node[black,midway, yshift = -.75cm]
            {$\displaystyle \Part{\textsf{n}_5}{2}$};
\path   (vertex=\n=1) ++ (-\factor*\outer, -\factor*\outer)
            coordinate (level 2 left) ;
\path   (vertex=\n=4) ++ (\factor*\outer, -\factor*\outer)
            coordinate (level 2 right) ;
\draw[decorate,decoration={brace,amplitude=10pt,mirror,raise=4pt},yshift=0pt]
    (level 2 left) -- (level 2 right)
        node[black,midway, yshift = -.75cm]
            {$\displaystyle \Part{\textsf{n}_5}{3}$};
\draw   (vertex=2=3) node [circle, draw=black, fill=level1color] {\phantom{5}};
\draw   (vertex=1=2) node [circle, draw=black, fill=level2color] {\phantom{5}};
\draw   (vertex=0=1) node [circle, draw=black, fill=level3color] {\phantom{5}};
\end{tikzpicture}}}
    \caption{$n_5$'s view of the network.}
    \label{fig:n5_view}
\end{figure}

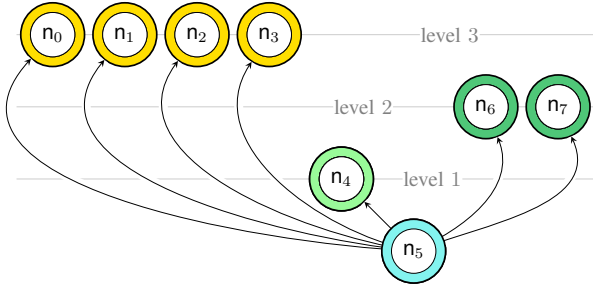
\begin{figure}[!ht]
    \centering
    \scalebox{0.80}{\trimbox{2cm 0cm 0cm 0cm}{\pgfmathsetmacro\vert{1.2}
\pgfmathsetmacro\hori{1.2}
\pgfmathsetmacro\outer{.53}
\pgfmathsetmacro\inner{21}
\pgfmathsetmacro\factor{.90}

\begin{tikzpicture}[every node/.style={inner sep=0,outer sep=0}]
\def\n{3}

\draw[color = \grisclair]
        (.5*\hori,0)
            -- node [circle, fill=white, pos=.75] {\color{\grisfonce}level $3$}
                (8.5*\hori,0);
\draw[color = \grisclair]
        (.5*\hori,-\vert)
            -- node [circle, fill=white, pos=.6] {\color{\grisfonce}level $2$}
                (8.5*\hori,-\vert);
\draw[color = \grisclair]
        (.5*\hori,-2*\vert)
            -- node [circle, fill=white, pos=.72] {\color{\grisfonce}level $1$}
                (8.5*\hori,-2*\vert);
        
\foreach \i in {0,...,\n} {
    \pgfmathsetmacro\TwoToTheN{2^\n}
    \foreach \j in {1,...,\TwoToTheN} {
        \path
            (\j*\hori,-\i*\vert)
                coordinate
                    (vertex=\i=\j);
    }
}

\draw[draw=black, thick, fill=nodecolor!50] (vertex=\n=6) circle(\outer);
\draw[draw=black, thick, fill=level1color] (vertex=2=5) circle(\outer);
%
\foreach \j in {1,...,4} {
    \draw[draw=black, thick, fill=level3color]
            (vertex=0=\j)
                circle (\outer)
                    node[inner sep = \outer, minimum size = \outer]
                        {};
    \draw[draw=black, thick, fill=level3color]
            (vertex=0=\j)
                node[circle, inner sep = .38cm]
                    (v=0=\j) {};
}
\foreach \j in {7,8} {
    \draw[draw=black, thick, fill=level2color]
            (vertex=1=\j)
                circle (\outer)
                    node[inner sep = \outer, minimum size = \outer]
                        {};
    \draw[draw=black, thick, fill=level2color]
            (vertex=1=\j)
                node[circle, inner sep = .38cm]
                    (v=1=\j) {};
}

\foreach \j in {5} {
    \draw[draw=black, thick, fill=level1color]
            (vertex=2=\j)
                circle (\outer)
                    node[inner sep = \outer, minimum size = \outer]
                        {};
    \draw[draw=black, thick, fill=level1color]
            (vertex=2=\j)
                node[circle, inner sep = .38cm]
                    (v=2=\j) {};
}

\def\angle{3}
\foreach \j in {1,...,4} {
    \pgfmathsetmacro\number{int(\j-1)}
    \draw (vertex=0=\j) node [circle, draw = black, fill = white, inner sep=0pt, minimum size = \inner] {$\textsf{n}_{\number}$};
    \draw[white, line width = .15cm]
            (vertex=3=6) 
                edge [in = 225 + \j*\angle, out = 180 - \j*\angle]
                    (v=0=\j);
}

\foreach \j in {1,...,4} {
    \pgfmathsetmacro\number{int(\j-1)}
    \draw (vertex=0=\j) node [circle, draw = black, fill = white, inner sep=0pt, minimum size = \inner] {$\textsf{n}_{\number}$};
    \draw[draw = black]
            (vertex=3=6) 
                edge [-stealth, in = 225 + \j*\angle, out = 180 - \j*\angle]
                    (v=0=\j);
}
\foreach \j in {7,8} {
    \pgfmathsetmacro\number{int(\j-1)}
    \draw (vertex=1=\j) node [circle, draw = black, fill = white, inner sep=0pt, minimum size = \inner] {$\textsf{n}_{\number}$};
    \draw[white, line width = .15cm]
            (vertex=3=6) 
                edge [in = -90 + \j*\angle, out = 45 - \j*\angle]
                    (v=1=\j);
}
\foreach \j in {7,8} {
    \pgfmathsetmacro\number{int(\j-1)}
    \draw (vertex=1=\j) node [circle, draw = black, fill = white, inner sep=0pt, minimum size = \inner] {$\textsf{n}_{\number}$};
    \draw[draw = black]
            (vertex=3=6) 
                edge [-stealth, in = -90 + \j*\angle, out = 45 - \j*\angle]
                    (v=1=\j);
}
\foreach \j in {5} {
    \pgfmathsetmacro\number{int(\j-1)}
    \draw (vertex=2=\j) node [circle, draw = black, fill = white, inner sep=0pt, minimum size = \inner] {$\textsf{n}_{\number}$};
    \draw[draw = white, line width = .15cm]
            (vertex=3=6) 
                --
                    (v=2=\j);
}
\foreach \j in {5} {
    \pgfmathsetmacro\number{int(\j-1)}
    \draw (vertex=2=\j) node [circle, draw = black, fill = white, inner sep=0pt, minimum size = \inner] {$\textsf{n}_{\number}$};
    \draw[draw = black]
            (vertex=3=6) 
                edge  [-stealth]
                    (v=2=\j);
}
\foreach \j in {7,8} {
    \pgfmathsetmacro\number{int(\j-1)}
    \draw (v=1=\j) node [circle, draw = black, fill = white, inner sep=0pt, minimum size = \inner] {$\textsf{n}_{\number}$};
}
\foreach \j in {5} {
    \pgfmathsetmacro\number{int(\j-1)}
    \draw (v=2=\j) node [circle, draw = black, fill = white, inner sep=0pt, minimum size = \inner] {$\textsf{n}_{\number}$};
}
\draw (vertex=2=5) node [circle, draw = black, fill = white, inner sep=0pt, minimum size = \inner] {$\textsf{n}_{4}$};
\draw[draw=black, thick, fill=nodecolor!50] (vertex=\n=6) circle(\outer);
\draw (vertex=3=6) node [circle, draw = black, fill = white, inner sep=0pt, minimum size = \inner] {$\textsf{n}_{5}$};
\end{tikzpicture}

}}
    \caption{$n_5$ communication organization.}
    \label{fig:n5_communication}
\end{figure}

Handel organizes a node's sequence
of aggregation steps using a binary tree. 
For instance, in figures~\ref{fig:n5_view}
and~\ref{fig:n5_communication}, 
both $\mathsf{n}_4$ and $\mathsf{n}_5$ aggregate contributions 
$c_4$ and $c_5$ at level 1, 
and  $\mathsf{n}_{4},\dots,\mathsf{n}_7$ aggregate $c_4,\dots,c_7$ at level 2. 

Formally, every node $i$ defines a partition $\;\Part{i}{0}\sqcup \Part{i}{1} \sqcup\cdots\sqcup \Part{i}{L}$ of the set of nodes in the network, where $L$ is the last level\footnote{ie. $L=\lceil \ln_2(N)\rceil$}: $\Part{i}{0}=\{i\}$ and inductively $\forall l\in \llbracket\,1, L\,\rrbracket$, $\sqcup_{0\leq j\leq l}\Part{i}{j} = \{$the nodes in the full binary subtree of height $l$ containing $i\}$.
We call $\Part{i}{l}$ node $i$'s \tb{level $l$ peer set}. See figure~\ref{fig:n5_view} showing how node $5$ partitions the set of nodes.

\subsection{Peer ranks and contribution selection}\label{sec:protocol:ranking}
We call \tb{Verification Priority} (VP) the ranks publicly given by a node to its peers.
The VP function is computable from public data---the node list and a seed shared by all nodes--- and acts as a pseudo random permutation $\vp_i(j)=\textit{rand}(N, \textit{seed}; i,j)\in\llbracket0, N-1\rrbracket$ of its last argument $j$, where $\textit{seed}$ is a seed value.

Nodes sort the received contributions by their sender's $\vp{}$, and select the ones in the scoring \tb{window}:
if $v$ denotes the highest $\vp_i{}$ among $i$'s incoming contributions, $i$ only \tb{score} contributions with  sender $\vp$ between $v$ and $v + \mathit{windowSize}$.

Let $c$ be the current best aggregate contribution at level $\ell$. The \tb{score} of an incoming level $\ell$ contribution $c'$ is, if $(c, c')$ is aggregable, the weight of the aggregate, otherwise, it is the weight of the contribution obtained by aggregating $c'$ with all previously verified and aggregable \textit{individual} level $\ell$ contributions.
The highest scoring contribution gets verified first. 

\colorlet{nodecolor}{yellow!50}

\begin{figure}[ht!]
\centering
\begin{tikzpicture}[
decoration={brace},
block/.style={
draw,
minimum size=6mm,
minimum height=7mm,
fill=white,
rectangle, 
align=center,
node distance=6mm,
font=\small},
score/.style={
draw,
minimum size=6mm,
minimum height=7mm,
fill=white,
rectangle, 
align=center,
node distance=7mm,
font=\small}
]

\node[block, draw] (n0) at (0,0) {0};
\node[score, draw, below of=n0] {};

\node[block, draw, right of=n0] (n1) {1};
\node[score, draw, below of=n1] {};

\node[block, draw, right of=n1, fill=nodecolor] (n2)  {2};
\node[score, draw, below of=n2, fill=nodecolor] (s2)  {4};

\node[block, draw, right of=n2] (n3)  {3};
\node[score, draw, below of=n3] {};

\node[block, draw, right of=n3, fill=nodecolor] (n4)  {4};
\node[score, draw, below of=n4, fill=nodecolor] {1};

\node[block, draw, right of=n4, fill=nodecolor] (n5)  {5};
\node[score, draw, below of=n5, fill=nodecolor] {6};

\node[block, draw, right of=n5, fill=nodecolor] (n6)  {6};
\node[score, draw, below of=n6, fill=nodecolor] {1};

\node[block, draw, right of=n6] (n7)  {7};
\node[score, draw, below of=n7] {};

\node[block, draw, right of=n7, fill=nodecolor] (n8)  {8};
\node[score, draw, below of=n8, fill=nodecolor] {7};

\node[block, draw, right of=n8] (n9)  {9};
\node[score, draw, below of=n9] {};

\draw[decorate,decoration={brace}]  (0.9,0.45) -- (3.3,0.45);
\node[font=\small] (n0) at (2.1,0.75) {window};

\draw[decorate,decoration={brace}]  (5.77,0.33) -- (5.77,-0.33);
\node[font=\small] (n0) at (6.12,0.018) {VP};

\draw[decorate,decoration={brace}]  (5.77,-0.39) -- (5.77,-1.04);
\node[font=\small] (n0) at (6.35,-.70) {scores};

\end{tikzpicture}
\caption{\tb{VP, window and scores}, with 5 contributions received from a total of 10 participants and a window size of 4. The contribution verified first is the contribution of VP $5$ with score $6$.}
\end{figure}
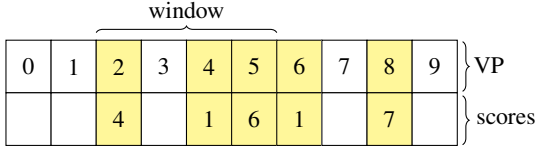

The result of the verification changes the window size. The size is increased by a factor two on verification success, decreased by a factor 4 on verification failure.

Nodes take into account other nodes' VP when they send their own contributions: inside a level, they send their contributions to those nodes whose VPs rank them highest. In other words, every node computes its level $\ell$ Contribution Prioritization Vector (CPV)
$[\vp_{s_1}(i),\dots,\vp_{s_{2^\ell}}(i)]$
where the $s_j$ are its level $\ell$ peers, and contacts the level $\ell$ peers that rank it highest, first.

\subsection{Initialization and node state}\label{sec:handel:state}
A Handel node receives from the application (1) the participants' public keys and network addresses, (2) a public random number. It creates the list of IDs by sorting the public keys, and shuffling them using the shared random number as a seed. It computes locally the levelwise $\vp_i$'s and CPV vectors.
The internal state of node $i$ consists of, for every level $\ell$:
\begin{compactenum}
    \item the $\vp_i(\ell)$ and CPV vectors,
    \item
    the current best incoming aggregate contribution $\inc_\ell^i$,
    \item
    the current best outgoing aggregate contribution $\out_\ell^i$ to send out,
    \item
    the list of as yet unverified contributions $\unv^i_{\ell,j}$ received from level $\ell$ peers $j$,
    \item
    the start time $\start{\ell}$,
    \item
    the current window size $\mathit{window}$.
\end{compactenum}
Let $\inc_0^i$ be $i$'s individual contribution. 
For all $\ell$, $\out_\ell^i=\mathit{Aggr}(\inc_0^i$,
$\inc_1^i,\dots{},\inc_{\ell-1}^i)$ is the
aggregate of the levelwise best received
contributions up to level $\ell-1$.
We call $\out_\ell^i$ (resp. $\inc_\ell^i$),
and level $\ell$ by extension, 
\textit{complete} if it includes 
all contributions from all levels
$\leq \ell$ (resp. $=\ell$).

\colorlet{inCol}{yellow!20}
\colorlet{outCol}{yellow!40}
\colorlet{finCol}{yellow!40}
\colorlet{sigCol}{white}

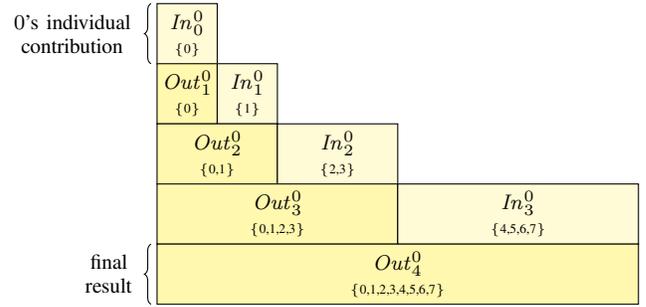
\begin{figure}[ht!]
\centering
\begin{tikzpicture}[
scale=0.80,
decoration={brace},
block/.style={
draw,
minimum size=10mm,
minimum height=10mm,
fill=white,
rectangle, 
align=center,
node distance=10mm,
font=\tiny}
]


\fill[fill=inCol, draw=black] (0,0) rectangle (1,-1) node[pos=.5, align=center, font=\footnotesize] {$\inc_0^0$\\\tiny\{0\}};

\fill[fill=outCol, draw=black]  (0,-1) rectangle (1,-2) node[pos=.5, align=center, font=\footnotesize] {$\out_1^0$\\\tiny \{0\}};
\fill[fill=inCol, draw=black] (1,-1) rectangle (2,-2) node[pos=.5, align=center, font=\footnotesize] {$\inc_1^0$\\\tiny\{1\}};

\fill[fill=outCol, draw=black] (0,-2) rectangle (2,-3) node[pos=.5, align=center, font=\footnotesize] {$\out_2^0$\\\tiny\{0,1\}};
\fill[fill=inCol, draw=black] (2,-2) rectangle (4,-3) node[pos=.5, align=center, font=\footnotesize] {$\inc_2^0$\\\tiny\{2,3\}};

\fill[fill=outCol, draw=black] (0,-3) rectangle (4,-4) node[pos=.5, align=center, font=\footnotesize] {$\out_3^0$\\\tiny\{0,1,2,3\}};
\fill[fill=inCol, draw=black] (4,-3) rectangle (8,-4) node[pos=.5, align=center, font=\footnotesize] {$\inc_3^0$\\\tiny\{4,5,6,7\}};

\fill[fill=finCol, draw=black] (0,-4) rectangle (8,-5) node[pos=.5, align=center, font=\footnotesize] {$\out_4^0$\\\tiny\{0,1,2,3,4,5,6,7\}};

\draw[decorate,decoration={brace,mirror}]  (-0.1,0) -- (-0.1,-1);
\node[align=center,font=\footnotesize] at (-1.4, -0.5) {$0$'s individual\\contribution};

\draw[decorate,decoration={brace,mirror}]  (-0.1,-4) -- (-0.1,-5);
\node[align=center,,font=\footnotesize] at (-0.8, -4.5) {final\\result};

\end{tikzpicture}
\caption{Example for \textbf{\textit{In} and \textit{Out} aggregate contributions} for node $0$, for aggregation with 8 nodes.}
\end{figure}

\subsection{Messages}
Handel uses a single message type for all communication. Messages contain a
\textit{level} $\ell$,
the \textit{sender's ID} $i$,
an \textit{aggregate contribution} $\out_{\ell}^i$ and the sender's \textit{individual contribution} $\inc_{0}^i$,
and a flag to signal that the receiver does not need to contact the sender. This flag is set when the sender has reached the threshold or when $\inc_{\ell}^i$ is complete.

\subsubsection{Sending contributions}
Two mechanisms trigger messages to be sent: \tb{periodic dissemination} and \tb{fast path}.
Periodic dissemination is triggered after every elapsed \textbf{dissemination period} (e.g. 20 milliseconds).
A node sends a message to one node from each of its \textit{active levels}.
Level $\ell$ is \textit{active} if $\out_{\ell}$ is complete or if $\start{\ell}$ has elapsed.
When a node completes $\out_{\ell}$ of some level $\ell$, the node triggers the fast path mode and sends $\out_{\ell}$ to a small number of peers (e.g. $10$) at this level $\ell$. 

\subsubsection{Receiving contributions}
When node $i$ receives a message from node $j$ at level $\ell$, it includes $j$'s aggregate and individual contributions ($\out_{\ell}^j$ and $\inc_{0}^j$ respectively) in its list of as-of-yet unverified level $\ell$ contributions.
The node continuously selects the best contribution to verify as described in section~\ref{sec:protocol:ranking}.

\subsubsection{Pruning contributions}
Handel \tb{prunes}
(1) all individual and aggregate contributions from nodes identified as Byzantine, e.g. which previously sent an invalid contribution,
(2) all individual and aggregate contributions at levels $\ell$ where $\inc_{\ell}$ is complete,
(3) all aggregate contributions that cannot be aggregated with, and have a lower score than $\inc_{\ell}$,
(4) all aggregate contributions from a given sender except that of highest weight.
This means that the contribution buffer size is bounded
: Handel keeps at most one contribution per participant.


\subsection{Parameter values}\label{sec:rationale}
We explained in section~\ref{sec:handel:presentation}
the rationale behind Handel's key mechanisms.
This section explains the parameter choices in light of associated trade-offs.
We designed Handel to perform optimally in a large-scale WAN and to require no further configuration regardless of node
capacity and connectivity.
To determine the optimal values of these parameters, we ran multiple simulations, with the parameters listed previously in section~\ref{sec:model:simulation}. We present here the graphs for 4096 nodes, with two configurations: one with 100\% of live nodes, one with 80\%.

\begin{figure}[ht!]
\begin{minipage}[t]{0.2\textwidth}
\begin{tikzpicture}
\begin{axis}[
scale=0.40,
ylabel=\textcolor{red}{---} Messages (thousands),
ymin=-1000,
ymax=11000,
ytick={0, 5000, 10000},
yticklabels={0,5,10},
xlabel=100\% live nodes]
\addplot[red,mark=*]
[discard if not={type}{30}
]table [x=analyzed, y=msg, col sep=comma] {csv_simulation/parameters.csv};  
\end{axis}
\begin{axis}[
scale=0.40,
ytick style={draw=none},
ymin=-100,
ymax=2100,
ytick={0, 1000, 2000},
yticklabels={},
ytick pos=right]
\addplot[blue,mark=*][
discard if not={type}{30}
] table [x=analyzed, y=time, col sep=comma] {csv_simulation/parameters.csv};
\end{axis}
\end{tikzpicture}
\end{minipage}
\begin{minipage}[t]{0.2\textwidth}
\begin{tikzpicture}
\begin{axis}[
scale=0.40,
ytick style={draw=none},
ymin=-1000,
ymax=11000,
ytick={0, 5000, 10000},
yticklabels={},
xlabel=80\% live nodes]
\addplot[red,mark=*]
[discard if not={type}{301}
]table [x=analyzed, y=msg, col sep=comma] {csv_simulation/parameters.csv};  
\end{axis}
\begin{axis}[
scale=0.40,
ylabel=\textcolor{blue}{---} Execution time (seconds),
ymin=-100,
ymax=2100,
ytick={0, 1000, 2000},
yticklabels={0, 1, 2},
ytick pos=right]
\addplot[blue,mark=*][
discard if not={type}{301}
] table [x=analyzed, y=time, col sep=comma] {csv_simulation/parameters.csv};
\end{axis}
\end{tikzpicture}
\end{minipage}
\caption{Impact of different settings for the \tb{dissemination period}, in milliseconds, on the execution time (blue line) and the number of messages (red line).}
\label{sim:dp}
\end{figure}
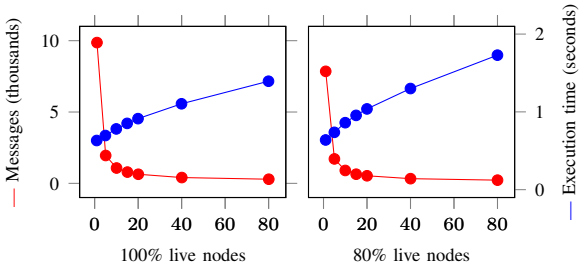

We chose a \tb{dissemination period} (DP) of 20 milliseconds. This allows multiple contributions to be verified within a single DP (e.g. BLS verifications take a few milliseconds), while being lower than  latencies typical in a WAN ($\approx$100 milliseconds).
Experimental results (figure \ref{sim:dp}) show this is a good trade-off between network bandwidth and time to completion.

The number of messages sent in \textbf{fast path} is set to 10.
The trade-off is between completion time bandwidth consumption.
Fast path is especially important when most nodes are available, as only completed levels can trigger it.
Fast path incurs no extra cost if it is never triggered.

\begin{figure}[ht!]
\begin{minipage}[t]{0.2\textwidth}
\begin{tikzpicture}
\begin{axis}[
scale=0.40,
ylabel=\textcolor{red}{---} Messages (hundreds),
ymin=400,
ymax=1100,
ytick={500, 750, 1000},
yticklabels={5,7.5,10},
xlabel=100\% live nodes]
\addplot[red,mark=*]
[discard if not={type}{20}
]table [x=analyzed, y=msg, col sep=comma] {csv_simulation/parameters.csv};  
\end{axis}
\begin{axis}[
scale=0.40,
ytick style={draw=none},
ymin=700,
ymax=1100,
ytick={800, 900, 1000},
yticklabels={},
ytick pos=right]
\addplot[blue,mark=*][
discard if not={type}{20}
] table [x=analyzed, y=time, col sep=comma] {csv_simulation/parameters.csv};
\end{axis}
\end{tikzpicture}
\end{minipage}
\begin{minipage}[t]{0.2\textwidth}
\begin{tikzpicture}
\begin{axis}[
scale=0.40,
ytick style={draw=none},
ymin=400,
ymax=1100,
ytick={500, 750, 1000},
yticklabels={},
xlabel=80\% live nodes]
\addplot[red,mark=*]
[discard if not={type}{201}
]table [x=analyzed, y=msg, col sep=comma] {csv_simulation/parameters.csv};  
\end{axis}
\begin{axis}[
scale=0.40,
ylabel=\textcolor{blue}{---} Execution time (seconds),
ymin=400,
ymax=1100,
ytick={500, 750, 1000},
yticklabels={0.8, 0.9, 1},
ytick pos=right]
\addplot[blue,mark=*][
discard if not={type}{201}
] table [x=analyzed, y=time, col sep=comma] {csv_simulation/parameters.csv};
\end{axis}
\end{tikzpicture}
\end{minipage}
\caption{Impact of different \tb{fast path} settings (number of nodes contacted) on execution time (blue line) and messages sent (red line).}
\label{sim:fp}
\end{figure}
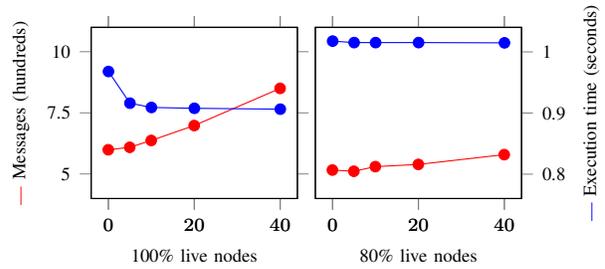

\tb{Level $\ell$ start time} is set to
$(\ell-1)\times50$ms.
Activating levels incrementally saves a significant number of messages when the protocol finishes quickly. Compared to a deployment without this technique  we observed similar times to completion but 20\% fewer messages.

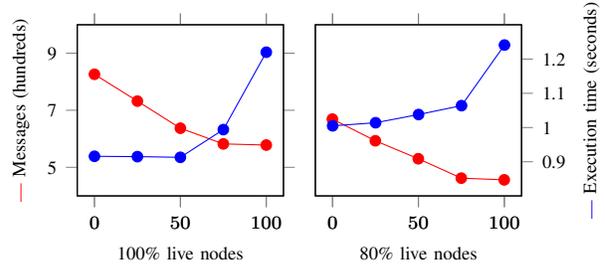
\begin{figure}[ht!]
\begin{minipage}[t]{0.2\textwidth}
\begin{tikzpicture}
\begin{axis}[
scale=0.40,
ylabel=\textcolor{red}{---} Messages (hundreds),
ymin=400,
ymax=1000,
ytick={500, 700, 900},
yticklabels={5,7,9},
xlabel=100\% live nodes]
\addplot[red,mark=*]
[discard if not={type}{10}
]table [x=analyzed, y=msg, col sep=comma] {csv_simulation/parameters.csv};  
\end{axis}
\begin{axis}[
scale=0.40,
ytick style={draw=none},
ymin=800,
ymax=1300,
ytick={900, 1000, 1100, 1200},
yticklabels={},
ytick pos=right]
\addplot[blue,mark=*][
discard if not={type}{10}
] table [x=analyzed, y=time, col sep=comma] {csv_simulation/parameters.csv};
\end{axis}
\end{tikzpicture}
\end{minipage}
\begin{minipage}[t]{0.2\textwidth}
\begin{tikzpicture}
\begin{axis}[
scale=0.40,
ytick style={draw=none},
ymin=400,
ymax=1000,
ytick={500, 700, 900},
yticklabels={},
xlabel=80\% live nodes]
\addplot[red,mark=*]
[discard if not={type}{101}
]table [x=analyzed, y=msg, col sep=comma] {csv_simulation/parameters.csv};  
\end{axis}
\begin{axis}[
scale=0.40,
ylabel=\textcolor{blue}{---} Execution time (seconds),
ymin=800,
ymax=1300,
ytick={900, 1000, 1100, 1200},
yticklabels={0.9, 1, 1.1, 1.2},
ytick pos=right]
\addplot[blue,mark=*][
discard if not={type}{101}
] table [x=analyzed, y=time, col sep=comma] {csv_simulation/parameters.csv};
\end{axis}
\end{tikzpicture}
\end{minipage}
\caption{Impact of different \tb{level start time} settings, in milliseconds, on execution time (blue line) and number of messages (red line).}
\label{sim:level-start-time}
\end{figure}

\tb{Windowing} preserves the benefits of scoring (verify only important contributions)  while ranking senders (which limits the effects of DoS attacks).
It is important that window size decrease quickly when under attack, so that Handel can rapidly return to using ranking alone.
Similarly, window size must grow quickly under normal conditions, so that nodes exploit incoming contributions to the fullest.
Window size expands and contracts exponentially. 
The \textbf{expansion factor} is $2$, while the \textbf{contraction factor} is $4$.
These values perform well under multiple attack scenarios, presented in section \ref{sec:attack scenarios}.
We limit the maximum window size to 128 to cap the time to return to a window size of 1 when attacked.

\section{Analysis}\label{sec:analysis}
We show in section~\ref{sec:security} that honest Handel nodes never generate invalid aggregate contributions and terminate under all circumstances.
In section~\ref{sec:attack scenarios} we 
analyze plausible attack scerarios.
In section~\ref{sec:complexity analysis} we briefly present our mathematical model of Handel and our main result, and use it to obtain Handel's
\tb{time complexity} (the time it takes all honest nodes to reach the threshold). From there, we derive Handel's
\tb{communication complexity} (the number of messages sent by an honest node) and
\textbf{processing complexity} (the number of verifications done by a node) in terms of the node count $N$.

\subsection{Safety and termination}\label{sec:security}

\tb{Safety: } 
\textit{Honest participants never produce invalid contributions.}
Indeed, honest participants always verify (1) the aggregability and (2) the validity of incoming contributions before aggregating them. Aggregation of valid, aggregable contributions produces valid aggregate contributions.

\noindent\tb{Termination: }
\textit{All honest participants eventually produce an aggregate signature of the required size.}
Since (1) the adversary cannot interfere with the contents of messages (2) the network guarantees message delivery, each honest participant eventually receives a message from every other honest participant (and, likewise, sends one to every honest participant). Messages include the sender's individual contribution; receivers do not prune these before including them in an aggregate contribution.

\subsection{Attack analysis and scenarios}\label{sec:attack scenarios}
While Byzantine nodes may behave arbitrarily, Handel only has one type of message and only one means of communication---pushing messages on authenticated channels---, thus greatly limiting an adversary's options.

The adversary can send \tb{many valid but useless contributions} to slow down the target or fill up its memory. Even so, memory consumption is limited and predictable: pruning of contributions means that Handel keeps at most one contribution per participant. Also, our assumption is that the cost of scoring is negligible (section~\ref{sec:model:contributions}). As a consequence we ignore this in our analysis\footnote{In practice, in an open network, it is always possible to interfere with the target's network and to perform a DoS on a participant. We discuss this in section~\ref{sec:anonymity}.}.
Since the adversary cannot prevent message delivery, honest nodes still receive contributions from other honest nodes.

The adversary can also refrain from participating, i.e. have its nodes behave like \tb{fail-silent} nodes. By neither verifying nor sending aggregate contributions to its peers, the adversary slows down the protocol. This is visible in figure~\ref{cpx-silent}.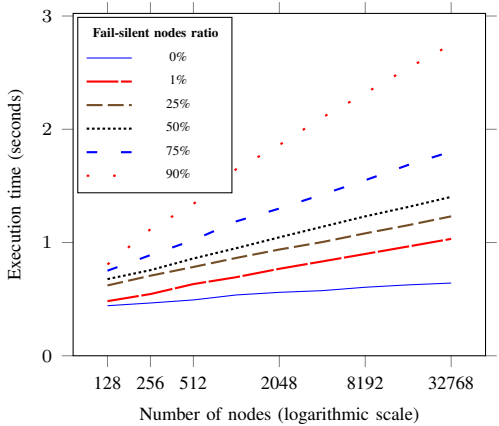
\begin{figure}[ht]
\begin{tikzpicture}
\begin{axis}[
xmode=log,
scale=0.8,
legend style={at={(0.01,0.99)},anchor=north west},
ymin=0,
y filter/.code={\pgfmathparse{#1/1000}\pgfmathresult},
xtick={128,256,512,2048,8192,32768},
xticklabels={128,256,512,2048,8192,32768},
ylabel=Execution time (seconds),
xlabel=Number of nodes (logarithmic scale)
]
\addlegendimage{empty legend}
\addlegendentry{\hspace{-.6cm}\textbf{Fail-silent nodes ratio}}
\addplot+[dashed1][
discard if not={dead}{0.0}
] table [x=node, y=time, col sep=comma] {csv_simulation/sim_all.csv};  
\addlegendentry{0\%}
\addplot+[dashed2][
discard if not={dead}{0.01}
] table [x=node, y=time, col sep=comma] {csv_simulation/sim_all.csv};  
\addlegendentry{1\%}
\addplot+[dashed3][
discard if not={dead}{0.25}
] table [x=node, y=time, col sep=comma] {csv_simulation/sim_all.csv};  
\addlegendentry{25\%}
\addplot+[dashed4][
discard if not={dead}{0.5}
] table [x=node, y=time, col sep=comma] {csv_simulation/sim_all.csv};  
\addlegendentry{50\%}
\addplot+[dashed5][
discard if not={dead}{0.75}
] table [x=node, y=time, col sep=comma] {csv_simulation/sim_all.csv};  
\addlegendentry{75\%}
\addplot+[dashed6][
discard if not={dead}{0.9}
] table [x=node, y=time, col sep=comma] {csv_simulation/sim_all.csv};  
\addlegendentry{90\%}
\end{axis}
\end{tikzpicture}
\caption{Aggregation time for different ratios of \tb{fail-silent nodes}, simulated with parameters~\ref{sec:model:simulation}.}
\label{cpx-silent}
\end{figure}

An adversary might also target an honest node by sending \tb{minimally interesting contributions} with high verification priority (VP) just in time to delay the verification and aggregation of large valid contributions with lower VP.
Since a node must send its individual contribution alongside an aggregate contribution, nodes sending minimal contributions will actually benefit the protocol more than nodes that don't participate at all. Furthermore, to move the honest contribution out of the scoring window, the adversary needs a high $\vp$.

\begin{figure}[ht]
\begin{tikzpicture}
\begin{axis}[
xmode=log,
scale=0.8,
legend style={at={(0.01,0.99)},anchor=north west},
ymin=0,
y filter/.code={\pgfmathparse{#1/1000}\pgfmathresult},
xtick={128,256,512,2048,8192,32768},
xticklabels={128,256,512,2048,8192,32768},
ylabel=Execution time (seconds),
xlabel=Number of nodes (logarithmic scale)]
\addlegendimage{empty legend}
\addlegendentry{\hspace{-.6cm}\textbf{Byzantine nodes ratio}}
\addplot+[dashed1][
discard if not={dead}{0.0}
] table [x=node, y=time, col sep=comma] {csv_simulation/sim_all.csv};
\addlegendentry{0\%}
\addplot+[dashed2][
discard if not={dead}{0.01}
] table [x=node, y=time, col sep=comma] {csv_simulation/sim_attack_hidden.csv};
\addlegendentry{1\%}
\addplot+[dashed3][
discard if not={dead}{0.25}
] table [x=node, y=time, col sep=comma] {csv_simulation/sim_attack_hidden.csv};
\addlegendentry{25\%}
\addplot+[dashed4][
discard if not={dead}{0.5}
] table [x=node, y=time, col sep=comma] {csv_simulation/sim_attack_hidden.csv};
\addlegendentry{50\%}
\addplot+[dashed5][
discard if not={dead}{0.75}
] table [x=node, y=time, col sep=comma] {csv_simulation/sim_attack_hidden.csv};
\addlegendentry{75\%}
\addplot+[dashed6][
discard if not={dead}{0.9}
] table [x=node, y=time, col sep=comma] {csv_simulation/sim_attack_hidden.csv};
\addlegendentry{90\%}
\end{axis}
\end{tikzpicture}
\caption{Aggregation time for different ratios of \tb{Byzantine nodes sending minimal contributions}.}\label{cpx-hidden}
\end{figure}
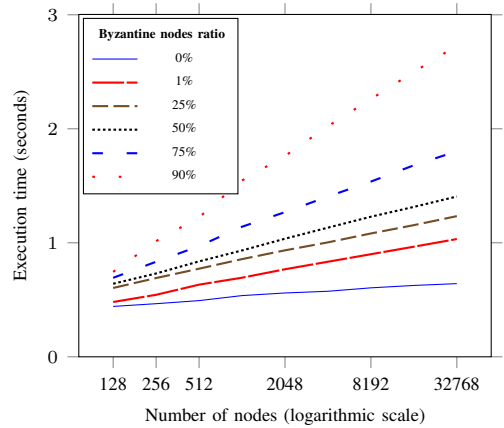

The last attack is sending \tb{high-scoring but invalid contributions} to delay the inclusion of other contributions.
In contrast to the previously described attacks, this attack is attributable. That is, it identifies the sender node as Byzantine.
In many circumstances, this makes the attack impractical.
For example, nodes identified as behaving in this way can be excluded from future aggregation rounds.
In a proof-of-stake crypto-currency context, the participant may lose (part of) its stake.
Nonetheless, we consider this attack to be the most damaging.
Indeed, verifying an invalid contribution constitutes a complete waste of the time and machine resources allocated to that task.

\begin{figure}[ht]
\begin{tikzpicture}
\begin{axis}[
xmode=log,
scale=0.8,
legend style={at={(0.01,0.99)},anchor=north west},
ymin=0,
y filter/.code={\pgfmathparse{#1/1000}\pgfmathresult},
xtick={128,256,512,2048,8192,32768},
xticklabels={128,256,512,2048,8192,32768},
ylabel=Execution time (seconds),
xlabel=Number of nodes (logarithmic scale)]
\addlegendimage{empty legend}
\addlegendentry{\hspace{-.6cm}\textbf{Byzantine nodes ratio}}
\addplot+[dashed1][
discard if not={dead}{0.0}
] table [x=node, y=time, col sep=comma] {csv_simulation/sim_all.csv};
\addlegendentry{0\%}
\addplot+[dashed2][
discard if not={dead}{0.01}
] table [x=node, y=time, col sep=comma] {csv_simulation/sim_attack_suicide.csv};
\addlegendentry{1\%}
\addplot+[dashed3][
discard if not={dead}{0.25}
] table [x=node, y=time, col sep=comma] {csv_simulation/sim_attack_suicide.csv};
\addlegendentry{25\%}
\addplot+[dashed4][
discard if not={dead}{0.5}
] table [x=node, y=time, col sep=comma] {csv_simulation/sim_attack_suicide.csv};
\addlegendentry{50\%}
\addplot+[dashed5][
discard if not={dead}{0.75}
] table [x=node, y=time, col sep=comma] {csv_simulation/sim_attack_suicide.csv};
\addlegendentry{75\%}
\addplot+[dashed6][
discard if not={dead}{0.9}
] table [x=node, y=time, col sep=comma] {csv_simulation/sim_attack_suicide.csv};
\addlegendentry{90\%}
\end{axis}
\end{tikzpicture}
\caption{Aggregation time for different ratios of \tb{Byzantine nodes sending invalid contributions}.}\label{cpx-suicide}
\end{figure}
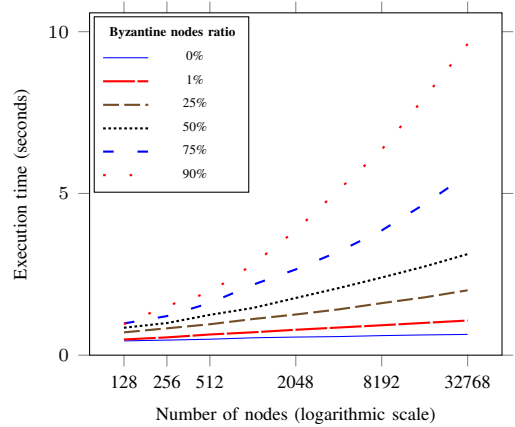

\subsection{Complexity analysis}\label{sec:complexity analysis}

\subsubsection{Handel convergence}
We consider in our mathematical model that (1) every participant has a unique ID. (2) Because of Handel's setup phase, Byzantine nodes are uniformly distributed in this ID space. (3) Nodes have uniformly drawn Verification Priorities (\vpl{}) on each level.
The mathematical model ignores some of Handel's features such as \textit{fast path} and \textit{scoring}, which increase performance in practice without fundamentally improving the time complexity.

The convergence theorem below expresses that, with overwhelming probability, all honest nodes will reach the threshold after sending and processing a polylogarithmic (in $N$) number of messages. A detailed mathematical model, along with a proof of the result, can be found in appendix~\ref{app: proof of convergence}.
\begin{thm}[Handel convergence]\label{thm:Handel Complexity, Byzantine case}
\emph{
Let $0<\byz_\mathrm{max}<1$ be an upper bound on the expected proportion of Byzantine nodes and let $0<\tau<1-\byz_\mathrm{max}$ be a target threshold.
There is a positive constant $C=C(\byz_\mathrm{max}, \tau)$ such that if honest nodes only read messages sent by the $\lceil C\ln(N)\rceil$ first nodes in every levelwise \vpl{}, and send at most $2\lceil C\ln(N)\rceil$ messages per level, then
\[
	\proba\left[
	\begin{array}{c}
	\text{all honest nodes aggre-}\\
	\text{gate a complete signature}
	\end{array}
	\right]
	\xrightarrow[\text{exp. fast}]{~\ln(N)\to+\infty~}1
\]
}
\end{thm}
We provide a sketch of proof. With overwhelming probability, Byzantine nodes are \textit{globally} homogeneously mixed up with honest nodes in ID space, see section~\ref{sec:handel:state}. This homogeneity actually already manifests \textit{locally}, i.e. within \textit{low level} peer sets.
By \textit{low level} we mean levels $\ell$ small enough for nodes to communicate with all their $\leq\ell$-level peers in $O(\ln(N))$ time.
This is precisely quantifiable using \textbf{concentration inequalities for binomial distributions}~\cite{concentrationChernoff, concentration_lecture_notes}. As a consequence, honest nodes find themselves with complete, threshold-meeting, level $\ell$-aggregate contributions. This pattern of homogeneous, threshold-meeting-contribution-owning collections of honest nodes is exceedingly likely to persist into higher levels with $O(\ln(N))$ messages sent, and verifications performed, per node and per level; this, too, follows from concentration inequalities. When this happens, all honest nodes find themselves with complete contributions at the final level.

\subsubsection{Time complexity}
Theorem~\ref{thm:Handel Complexity, Byzantine case} allows us to analyze a more stringent condition than all 
honest nodes meeting the threshold: that of \textit{all 
honest nodes aggregating a complete contribution}. When the proportion of honest nodes is greater than the threshold ($\tau<1-b_{\mathrm{max}}$), complete contributions meet the threshold.

Let $\mathsf{Lat}_\mathrm{max}$ denote the maximal network latency and let $\Delta\mathsf{Start}_\mathrm{max}$ denote the maximal delay between Handel nodes starting the protocol (recall our assumptions on participants
\ref{sec:model, participants}). Let $\mathsf{Ver}_\mathrm{max}$ denote the maximal contribution verification time
and let $\mathsf{DP}$ denote the dissemination period. It takes up to $2\lceil C\ln(N)\rceil\cdot\mathsf{DP} + \mathsf{Lat}_\mathrm{max}$ for all messages sent by a node at some level to reach their destination, and up to $\lceil C\ln(N)\rceil\cdot\mathsf{Ver}_\mathrm{max}$ and for a node to verify the expected $\lceil C\ln(N)\rceil$  (an honest recipient may receive all $\lceil C\ln(N)\rceil$ expected incoming contributions from the initial segment of its levelwise VP at the same time). It may happen that nodes make progress one level at a time (instead of progressing in parallel on all levels); since there are $\lceil\ln_2(N)\rceil$ levels, this puts the maximal time complexity per node for a Handel run at
\[
    \Delta\mathsf{Start}_\mathrm{max}
    +
    \lceil\ln_2(N)\rceil\cdot
    \left(
    \begin{array}{l}
        2\lceil C\ln(N)\rceil\cdot\mathsf{DP} + \mathsf{Lat}_\mathrm{max}\\[1mm]
        ~+ \lceil C\ln(N)\rceil\cdot\mathsf{Ver}_\mathrm{max} 
    \end{array}
    \right)
\]
This analysis puts time complexity at $O(\ln(N)^2)$, with nodes sending a total of $\lceil\ln_2(N)\rceil\cdot 2\lceil C\ln(N)\rceil=O(\ln(N)^2)$ messages and verifying up to $\lceil\ln_2(N)\rceil\cdot\lceil C\ln(N)\rceil=O(\ln(N)^2)$ contributions.

\subsubsection{Conclusion}
Handel's \tb{time complexity} is thus $O(\ln(N)^2)$. Its \tb{message complexity} is actually $O(\ln(N)^3)$, since Handel sends up to $\ln(N)$ messages \textit{every} dissemination period. However, because of our result on time complexity, the \tb{processing complexity}, i.e. the number of verified messages per node is $O(\ln(N)^2)$.

\section{Anonymity}\label{sec:anonymity}
Handel requires participants to reveal their IP addresses in order to receive Handel packets.
In some cases, the mapping between public keys and IP addresses being public is problematic as it leaves signers vulnerable to targeted DoS attacks and privacy breaches.
For instance, in proof-of-stake protocols, a signer may want to hide its IP address if it is linked with the public key controlling its staked crypto-currency funds.
We first consider the ``vanilla'' deployment of Handel where IP addresses are public.
The second deployment option uses an anonymity network (such as Tor~\cite{dingledinetor}) and its synergies with Handel.
Thirdly, we describe a deployment scenario where ring signatures provide signers anonymity within the set of participants.

\subsection{``Vanilla'' deployment} 
In this deployment, every node is reachable from the Internet. There is a public global directory containing the mapping between signers' public keys and their public addresses. This deployment scenario is standard in distributed systems and, as such, well understood. However, having signers' IP addresses be public exposes them to DoS attacks or even full compromise.
%

There is no simple solution against full compromise. A signer can put its signing key in an external HSM module~\cite{cifuentes2016poor}. Such modules typically include rate limiters, so lowers the impact of a successful hack. Another argument is Handel can be put on a separate node than the application that creates the initial contribution. Handel will hide the real machine, while its limited scope and communication type will make it simpler to protect against hacking.

DoS attacks can either target a small number of participants, with the objective of preventing them from having their contributions included, or target a large number of participants, with the objective of slowing down aggregation or making it impossible to reach the threshold.
DoS-ing a single participant is possible, with the caveat that one packet reaching one honest node is enough to cancel most of the attack: this honest node will include the contribution in its own aggregate contributions, and will thus disseminate it to others.
In addition,
since Handel eschews any timeout-based mechanism (section~\ref{sec:handel:presentation}) means that attackers cannot make honest nodes seem dead/Byzantine to others.
The feasability of attacking the protocol, preventing it from reaching the threshold, depends on the size of the aggregator set.
The whole point of Handel is to allow very large aggregator sets for this exact reason.
For instance, with an aggregator set of 4000 nodes, DoS-ing 1000 nodes represents 25\% of the network, yet the impact on Handel's execution time is minimal, as seen in figure~\ref{fig:failure:time}.
Lastly, a larger fraction of Byzantine nodes means slower aggregation, but the exponents in the asymptotic bounds do not change. Handel does not have ``critical thresholds'' that the attacker could target: there is no inherent \textit{qualitative} difference between $33\%$ and $34\%$ or between $49\%$ and $51\%$ of the nodes not participating.

\subsection{Anonymity network}
A participant wishing to hide its public IP address can use Tor~\cite{dingledinetor}, I2P~\cite{i2p} or Freenet~\cite{clarke1999distributed} to deploy hidden services and use the service address instead of a public IP address. In this deployment, \ti{anonymity guarantees} are far greater than in the vanilla deployment. In order to DoS a signer, an attacker would have to either attack the anonymity
network itself, or de-anonymize the signer. While the latter is possible for state-level adversaries, it is a nontrivial attack. Regarding the former, Tor has implemented multiple DoS protections~\cite{torDoSDefense} and the most effective DoS attack remains that against the whole Tor network.

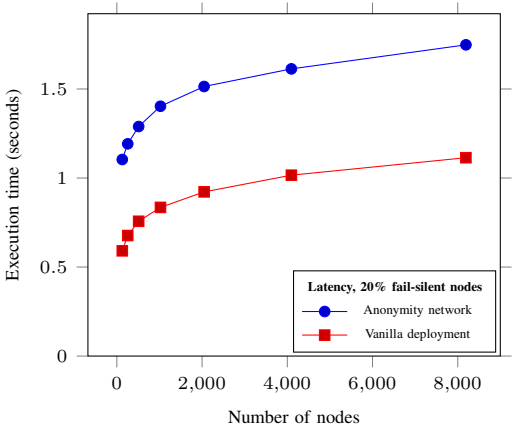
\begin{figure}[ht]
\begin{tikzpicture}
\begin{axis}[
legend style={at={(0.99,0.01)},anchor=south east},
scale=0.8,
ymin=0,
y filter/.code={\pgfmathparse{#1/1000}\pgfmathresult},
ylabel=Execution time (seconds),
xlabel=Number of nodes
]
\addlegendimage{empty legend}
\addlegendentry{\hspace{-.6cm}\textbf{Latency, 20\% fail-silent nodes}}

\addplot+[][discard if not={tor}{0.2}
] table [x=node, y=time, col sep=comma] {csv_simulation/tor.csv};  
\addlegendentry{Anonymity network}

\addplot+[][discard if not={tor}{0}
] table [x=node, y=time, col sep=comma] {csv_simulation/tor.csv};  
\addlegendentry{Vanilla deployment}

\end{axis}
\end{tikzpicture}
\caption{Execution time with 20\% fail-silent nodes and 20\% of nodes having an extra \tb{round-trip latency of 1000ms} compared to a vanilla deployment with location-based latencies only.}
\label{fig:comparative:tor}
\end{figure}

In practice, Handel participants could be on different networks, e.g. some on Tor, others using their real IP address.
A few participants using Tor will not impact either Tor or Handel, but Handel is designed for large scale aggregation.
For instance, a deployment with 8000 Handel nodes, 20\% of which use Tor, represents 1600 nodes sending dozens of messages per second over Tor.
While we do not believe such a deployment to be practical today on Tor, we can nevertheless analyze it.
To do so, we must take into account the impact using an anonymity network has on latency: Tor messages typically take upwards of 500 milliseconds~\cite{torPerf} to arrive, while UDP messages typically reach their destination within 100-200 milliseconds~\cite{latencyInternet}.
We simulated a configuration with 20\% of nodes having an extra round-trip latency of 1000ms, and with 20\% fail-silent nodes overall. The results (figure~\ref{fig:comparative:tor}) show that Handel reaches the 79.9\% threshold in 1.7 seconds compared to 1.1 seconds in a vanilla deployment.

\subsection{Anonymous signers}
In the vanilla deployment, network addresses and the mapping between these and public keys are public. In the anonymity network deployment, the real network addresses are hidden.
In the anonymous signers deployment, described here, new keys are created from the set of original keys, but it is impossible to link any new key to any of the the original public keys. These new keys are linked to network addresses.
Hence it is not possible to link the network address to the original public key.

\newcommand\xsetpos{2}
\newcommand\xsetposs{5}
\newcommand\xsetposss{7}
\newcommand\xsetpossss{9}

\begin{figure}[ht!]
    \centering
\begin{tikzpicture}[scale=.7,
                    arrow/.style={thick,->,black},
                    set name/.style={font=\color{black}\Large\bfseries\sf},
                    set/.style={black},
                    every node/.style={circle},
                    font=\footnotesize
                    ]
\draw[set] (0,0) circle [x radius=0.75cm, y radius=1.7cm]
           (\xsetpos,0) circle [x radius=0.75cm, y radius=1.7cm]
           (\xsetposs,0) circle [x radius=0.75cm, y radius=1.7cm]
           (\xsetposss,0) circle [x radius=0.75cm, y radius=1.7cm]
           (\xsetpossss,0) circle [x radius=0.75cm, y radius=1.7cm];

\node[] (s1) at (5.54,0) {};

\node[set name, align=center] at (0,-2)  {\tiny public keys};
\node[set name, align=center] at (\xsetpos,-2) {\tiny network addresses};  
\node[set name, align=center] at (\xsetposs,-2) {\tiny public keys}; 
\node[set name, align=center] at (\xsetposss,-2) {\tiny new public keys};
\node[set name, align=center] at (\xsetpossss,-2) {\tiny network addresses};  

\node[set name, align=center] at (1,-2.8) {\small Vanilla};
\node[set name, align=center] at (\xsetposss,-2.8) {\small Anonymous signers};  

\node[inner sep=0] (a1) at (0,1)  {$Pk_1$};
\node[inner sep=0] (a2) at (0,0)  {$Pk_2$};
\node[inner sep=0] (a3) at (0,-1) {$Pk_n$};

\node[inner sep=0] (b1) at (\xsetpos,1)  {$N_1$};
\node[inner sep=0] (b2) at (\xsetpos,0)  {$N_2$};
\node[inner sep=0] (b3) at (\xsetpos,-1) {$N_n$}; 

\begin{scope}[<->]
  \draw (a1.east) -- (b1);
  \draw (a2.east) -- (b2);
  \draw (a3.east) -- (b3);
\end{scope}

\node[inner sep=0] (c1) at (\xsetposs,1)  {$Pk_1$};
\node[inner sep=0] (c2) at (\xsetposs,0)  {$Pk_2$};
\node[inner sep=0] (c3) at (\xsetposs,-1) {$Pk_n$};

\node[inner sep=0] (d1) at (\xsetposss,1)  {$AK_1$};
\node[inner sep=0] (d2) at (\xsetposss,0)  {$AK_2$};
\node[inner sep=0] (d3) at (\xsetposss,-1) {$AK_n$}; 

\node[inner sep=0] (e1) at (\xsetpossss,1)  {$N_1$};
\node[inner sep=0] (e2) at (\xsetpossss,0)  {$N_2$};
\node[inner sep=0] (e3) at (\xsetpossss,-1) {$N_n$}; 

\begin{scope}[<->]
  \draw (s1) -- (d1);
  \draw (s1) -- (d2);
  \draw (s1) -- (d3);
\end{scope}

\begin{scope}[<->]
  \draw (d1.east) -- (e1);
  \draw (d2.east) -- (e2);
  \draw (d3.east) -- (e3);
\end{scope}
\end{tikzpicture}    
\vspace*{-15mm}
\caption{\tb{Anonymous signers vs. Vanilla deployment}. There is exactly one new public key for each public key in the original set, but it is impossible to establish the link between the two keys.}
\label{fig:deps}
\end{figure}
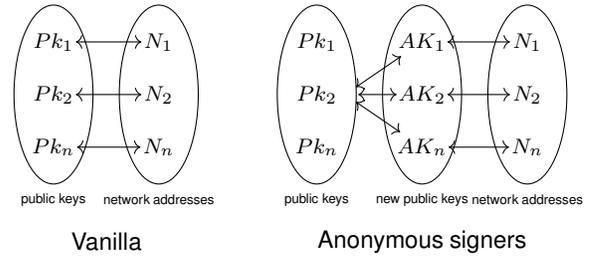

This can be achieved if participants are allowed to sign messages using fresh secret/public key pairs.

By means of \ti{linkable ring signatures}~\cite{liu2005linkable}, individual public keys $PK_i$ from a set of public keys $PK=\{PK_i\}_i$ can create new ``anonymous'' public keys $AK_i$ in such a way that any attempt to create multiple $AK_i$ from a single public key $PK_i$ is apparent.

Using anonymous public keys in Handel requires a setup phase where every signer $i$ produces a new anonymous private / public key pair $ak_i\;/\:AK_i$.

\noindent\textbf{Setup phase.}
Every signer locally creates a new private key $ak_i$  and public key $AK_i$. It then broadcasts the message  $\textit{IP-address}_i\|AK_i$ with $\sigma_i$, a linkable ring signature for the set of signers $PK$  of this message. At the end of the setup phase, the signers know the list of IP addresses of the other nodes and their new anonymous public keys $AK_i$. If an honest signer detects two signatures from the same signer (through \ti{linkability}), the associated $AK_i$ is discarded from $AK$. Using \ti{traceable ring signatures}~\cite{fujisaki2008traceable}, the owner of $PK_i$ can even be identified and removed from the public list of signers. 

Handel then runs its course using the new set of public keys $AK$, without the mapping between signing keys and IP addresses.

The anonymity protection offered by this protocol increases as more signers opt in. This is fortunate considering Handel's ability to scale. Importantly, the setup phase need only be carried out \textit{once} and can be used for multiple aggregation rounds.

\section{Implementation}\label{sec:impl}
We released a reference implementation of Handel as an open source Golang library~\cite{handellib} under the Apache license. The library implements the protocol described in section~\ref{sec:protocol} in $\sim$3000 lines of code. It is able to accommodate different \textit{aggregation schemes} (BLS with BN256, BLS12-381, etc) and \textit{network layers} (UDP, QUIC, etc.) We implemented BLS aggregation on the BN256~\cite{naehrig2010new} curve (implementation by Cloudflare~\cite{bn256cf}), with UDP as the transport layer. Our implementation includes a few technical optimizations. Specifically, in QUIC it is possible to know if a sent message was received. Therefore, when using QUIC, we do not send the same message twice to a  peer. As an alternative, nodes using raw UDP include a flag in their messages to signal they already received a message, so nodes do not keep sending messages in the lower levels.

\tb{Isolated verification.}
Handel strives to minimize the number of CPU cores dedicated to scoring and contribution verification.
Since Handel is to be used alongside an application, Handel's CPU consumption must neither impact nor hinder the application's normal functioning. Our implementation thus  scores and verifies incoming contributions on a single CPU core.

\tb{Message format.}
Messages in the reference implementation contain: the sender's ID (a 32 bit integer), the level (a byte), an aggregate contribution (as defined in section~\ref{sec:model:contributions}) and the signature of the sender (i.e. its individual contribution).
A BN256 signature is 64 bytes. The size of the bitset depends on the level
($\textit{size}=2^{\textit{level} -1}$ bits).
For 4000 nodes, the bitset size can go up to 250 bytes.
The maximum total message size is thus about 400 bytes. 

\tb{Wire protocol.} We used the gob encoding scheme~\cite{gobGolang} which adds a few extra bytes to messages.

\section{Evaluation}\label{sec:eval}
We present our test results for multi-signature aggregation using Handel. Our focus is on Handel's performance and comparison with other aggregation protocols. We observe logarithmic completion times and resource consumption.

\subsection{Experimental setup}\label{sec:eval-setup}

\noindent\tb{Measurements.}
All Handel nodes log measurements such as the total number of messages received/sent, verifications performed, and other internal statistics.

\noindent\tb{Network topology.}
We ran large scale experiments of Handel on AWS EC2 instances.
Our biggest tests involved  2000 t2.nano instances, each with a 3.3Ghz core.
In most cases, we ran two Handel nodes per EC2 instance, sometimes incurring an over-subscription effect.
Therefore, we expect real-world deployments on commodity hardware to exhibit better performance than those measured in our tests.
In order to simulate a real world deployment, and latency variability in particular, we used AWS instances from 10 AWS regions located all over the world. The measured latencies are presented in table~\ref{tab:latencies}. Regions span the entirety of the globe and therefore give rise to variations of more than 250 milliseconds.

\begin{table*}
\small
\centering
    \begin{tabular}{|c||*{10}{c|}}
        \hline
        Regions & Virginia & Mumbai & Seoul & Singapore & Sydney & Tokyo & Canada & Frankfurt & Ireland & London \\
        \hline\hline
        Oregon &  81 & 216 & 126 & 165 & 138 & 97 & 64 & 164 & 131 & 141 \\
        Virginia & - & 182 & 181 & 232 & 195 & 167 & 13 & 88 & 80 & 75 \\
        Mumbai & - & - & 152 & 62 & 223 & 123 & 194 & 111 & 122 & 113 \\
        Seoul  & - & - & - & 97 & 133 & 35 & 184 & 259 & 254 & 264 \\
        Singapore  & - & - & - & - & 169 & 69 & 218 & 162 & 174 & 171 \\
        Sydney  & - & - & - & - & - & 105 & 210 & 282 & 269 & 271 \\
        Tokyo  & - & - & - & - & - & - & 156 & 235 & 222 & 234 \\
        Canada  & - & - & - & - & - & - & - & 101 & 78 & 87 \\
        Frankfurt  & - & - & - & - & - & - & - & - & 24 & 13 \\
        Ireland  & - & - & - & - & - & - & - & - & - & 12 \\
        \hline
    \end{tabular}
    \vspace*{1mm}
    \caption{\tb{AWS latencies}, in milliseconds. Latencies are assumed to be symmetrical.}
    \label{tab:latencies}
\end{table*}

\subsection{Comparative baseline}\label{sec:eval:baseline}

\begin{figure}[ht]
\begin{tikzpicture}
\begin{axis}[
legend style={at={(0.01,0.99)},anchor=north west},
scale=0.8,
xtick={100,1000,2000,3000,4000},
xticklabels={100,1000,2000,3000,4000},
ylabel=Execution time (seconds),
xlabel=Number of nodes
]
\addlegendimage{empty legend}
\addlegendentry{\hspace{-.6cm}\textbf{Protocol}}

\addplot+[
] table [x=totalNbOfNodes, y=sigen_wall_avg, col sep=comma] {csv_evaluation/handel_0failing_99thr.csv};  
\addlegendentry{Handel}

\addplot+[
] table [x=totalNbOfNodes, y=sigen_wall_avg, col sep=comma] {csv_evaluation/n2_4000_99thr.csv};  
\addlegendentry{Complete}

\end{axis}
\end{tikzpicture}
\caption{\tb{Execution time} comparison up to 4000 nodes with a 99\% threshold}
\label{fig:comparative:time}
\end{figure}
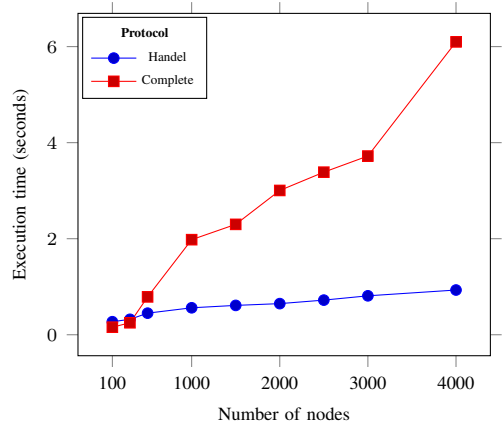

\begin{figure}[ht]
\begin{tikzpicture}
\begin{axis}[
ymode=log,
legend style={at={(0.01,0.99)},anchor=north west},
scale=0.8,
ytick={10240,102400,1048576,10485760},
yticklabels={10KB,100KB,1MB,10MB},
xtick={100,1000,2000,3000,4000},
xticklabels={100,1000,2000,3000,4000},
ylabel=Network data sent (logarithmic scale),
xlabel=Number of nodes
]
\addlegendimage{empty legend}
\addlegendentry{\hspace{-.6cm}\textbf{Protocol}}

\addplot+[
] table [x=totalNbOfNodes, y=net_sentBytes_avg, col sep=comma] {csv_evaluation/handel_0failing_99thr.csv};  
\addlegendentry{Handel}

\addplot+[
] table [x=totalNbOfNodes, y=net_sentBytes_avg, col sep=comma] {csv_evaluation/n2_4000_99thr.csv};  
\addlegendentry{Complete}

\end{axis}
\end{tikzpicture}
\caption{\tb{Outgoing network data} comparison up to 4000 nodes with a 99\% threshold}
\label{fig:comparative:network}
\end{figure}
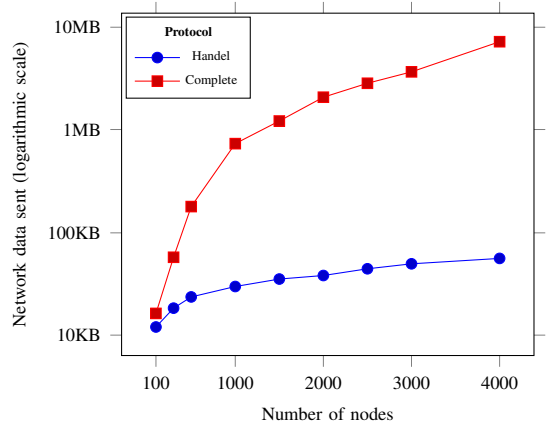

We compared Handel against a ``complete graph'' aggregation scenario.
In the complete graph scenario, nodes send their individual signatures to all other nodes.
In both scenarios, completion means aggregating 99\% of all individual signatures.
See figure~\ref{fig:comparative:time} for a comparison of completion times and figure~\ref{fig:comparative:network} for a comparison of the outgoing network consumption.

We observe that, on average, Handel is able to aggregate 4000 signatures in under a second.
Furthermore, Handel is both orders of magnitude faster and more resource efficient than the complete graph approach.
Finally, Handel's polylogarithmic (empirically logarithmic) time complexity is apparent from  figure~\ref{fig:comparative:time}. 

\subsection{Robustness}\label{sec:eval:failures}
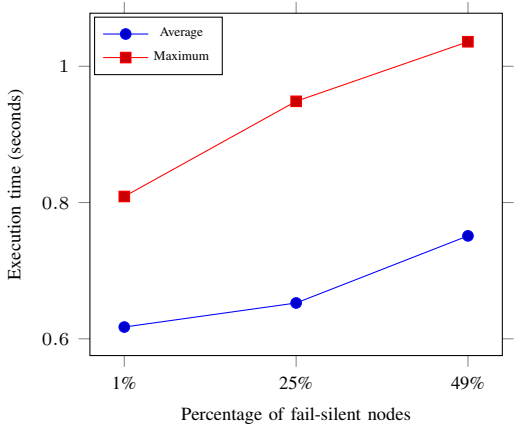
\begin{figure}[ht]
\begin{tikzpicture}
\begin{axis}[
legend style={at={(0.01,0.99)},anchor=north west},
scale=0.8,
xtick={40,1000,1960},
xticklabels={1\%,25\%,49\%},
ylabel=Execution time (seconds),
xlabel=Percentage of fail-silent nodes
]

\addplot+[
] table [x=failing, y=sigen_wall_avg, col sep=comma] {csv_evaluation/handel_4000_failing.csv};  
\addlegendentry{Average}

\addplot+[
] table [x=failing, y=sigen_wall_max, col sep=comma] {csv_evaluation/handel_4000_failing.csv};  
\addlegendentry{Maximum}

\end{axis}
\end{tikzpicture}
\caption{Various percentages of \tb{failing nodes} over a total of 4000 nodes for a 51\% threshold}
\label{fig:failure:time}
\end{figure}

In these experiments, we tested Handel's resistance to node failures. Figure~\ref{fig:failure:time}
illustrates Handel's ability to aggregate 51\% of signatures from 4000 nodes at various rates of node failure.

\subsection{CPU consumption}\label{sec:eval:cpu}
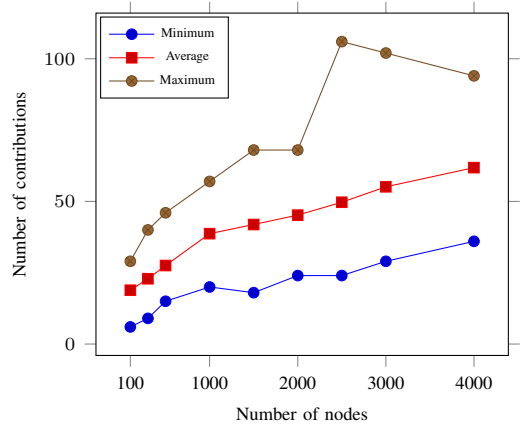
\begin{figure}[ht!]
\begin{tikzpicture}
\begin{axis}[
legend style={at={(0.01,0.99)},anchor=north west},
scale=0.8,
xtick={100,1000,2000,3000,4000},
xticklabels={100,1000,2000,3000,4000},
ylabel=Number of contributions,
xlabel=Number of nodes
]

\addplot+[
] table [x=totalNbOfNodes, y=sigs_sigCheckedCt_min, col sep=comma] {csv_evaluation/handel_0failing_99thr.csv};  
\addlegendentry{Minimum}

\addplot+[
] table [x=totalNbOfNodes, y=sigs_sigCheckedCt_avg, col sep=comma] {csv_evaluation/handel_0failing_99thr.csv};  
\addlegendentry{Average}

\addplot+[
] table [x=totalNbOfNodes, y=sigs_sigCheckedCt_max, col sep=comma] {csv_evaluation/handel_0failing_99thr.csv};  
\addlegendentry{Maximum}

\end{axis}
\end{tikzpicture}
\caption{Number of \tb{contributions checked} by a node, with up to 4000 total nodes and a 99\% threshold}
\label{fgi:cpu}
\end{figure}

This experiment was designed to test Handel's CPU
consumption under heavy load.
In Handel, signature verification is the main CPU bottleneck.
Figure~\ref{fgi:cpu} shows the \ti{\{minimum, maximum, average\}} number of signatures checked for different node counts.
All curves exhibit the expected polylogarithmic behaviour.
In particular, the minimum curve shows that some nodes verified close to the least possible number of aggregate signatures
for 4000 nodes: around $30$ vs. a theoretical minimum of $24= 2\cdot\lceil\ln_2(4000)\rceil$ (recall that every message
contains \textit{two} signatures).
The graph also shows that some nodes verified \ti{far more} signatures than that.
High latencies contribute to this discrepancy: some nodes may have to wait \ti{longer} to receive interesting contributions.
In the meantime, they will verify all other incoming signatures, eventually even those that score poorly.

\section{Related work}
The growing need for scalable distributed data aggregation systems has spawned a number of interesting designs in the literature.
These systems are primarily designed for \textit{permissioned networks}  such as grid networks or monitoring networks.

\subsection{\sanFermin{}}\label{sec:rel:sanfermin}
\sanFermin{}~\cite{cappos2008san} introduced the \ti{binomial swap forest} technique enabling $O(log(N))$ time complexity for its aggregation protocol when there is no failure. \sanFermin{} is defined in the \ti{fail-silent} model where a node can fail and stop responding. Byzantine actors can prevent a contribution from being included by contacting high level nodes early on. Moreover, failure to send one's contribution within a limited amount of time is interpreted as node failure. Hence, timeouts are a key parameter of the system: if timeouts are too short, nodes might be evicted too early; long timeouts on the other hand negatively impact time to completion. While Handel reuses \sanFermin{}'s overlay structure, allowing for Byzantine nodes, and forgoing timeouts leads to major differences from \sanFermin{}: (1) Handel nodes send \textit{one-way} messages to \textit{multiple} nodes on \textit{all} levels concurrently, and select the best incoming contributions for aggregation. At any point in time, \sanFermin{} does \textit{two-way} communication with a \textit{single} node at a \textit{single} level, and uses a timeout to decide that a node has failed. (2) In Handel all levels are activated in parallel, and aggregate contributions are updated. In \sanFermin{}, once a node has successfully contacted another, it moves to the next level, whatever the size of the contribution it received. (3) In Handel, contributions are scored, allowing slow nodes (eg. weak machines or nodes flooded with Byzantine contributions) to optimize resource usage. There is no comparable mechanism in \sanFermin{}. (4) Handel uses prioritizing/windowing to protect against Byzantine attacks targeted at wasting participants' resources.
Again, there is no comparable mechanism in \sanFermin{}.

This leads to practical differences: Handel is Byzantine-tolerant and can be deployed on a WAN with heterogeneous nodes. The impact of network latency, node failures and slow nodes on Handel is limited, while under these conditions \sanFermin{}'s completion time will depend on the timeout.

\subsection{Gossip-based solutions}
The systems from \cite{gupta2001scalable,kempe2003gossip,jelasity2005gossip} are gossip-based aggregation protocols. 
Gupta et al.~\cite{gupta2001scalable} use a tree-based hierarchy protocol where participants cluster in groups to form the nodes ("boxes") of the tree. The protocol works in phases where participants in a given "box" aggregate their values via gossiping, then pass the aggregated value to the parent "box" in the next phase. Under low network failure and churn, the protocol exhibits a near-optimal time complexity of $O(\ln(N)^2)$. However, the protocol is not resistant against arbitrary Byzantine failures since the hierarchical structure means that a single corrupted "box" can eclipse the contributions from the nodes in the subtree beneath it.
Jelasity et al.~\cite{jelasity2005gossip} offer an \textit{adaptive} aggregation protocol that continuously tracks changes in input values reasonably quickly. The protocol uses a "push-pull gossip" based approach where each participant periodically and \textit{randomly} contacts new nodes in the network to update its internal state. This protocol is geared towards tasks such as continuous evaluation of a grid network rather than one shot queries where the input changes each time.
Neither solution is Byzantine-tolerant and both have greater aggregation time than Handel even with an order of magnitude fewer nodes. 

\subsection{Tree-based solutions}
Many systems use tree-based constructions to address scalability.
Astrolab~\cite{van2003astrolabe} is the first system using hierarchy zones, mimicking the design of DNS zones. Systems such as SDIMS~\cite{yalagandula2004sdims}, CONE~\cite{conedht}, and Willow~\cite{van2004willow} combine hierarchy-based constructions with the design principles underlying \ti{distributed hash table} (DHT) systems.
Li's protocol~\cite{li2005implementing} is natively compatible with any DHT abstraction.
However, individual nodes are solely responsible for aggregating and forwarding contributions from their respective subtrees; a single node failure can thus severely impact the protocol.

Finally, Grumbach et al.~\cite{grumbachdht} use a mechanism similar to binomial swap forests with the native underlying trees of the Kademlia DHT~\cite{maymounkov2002kademlia}.
In particular, it is the first system tackling aggregation with \ti{some} Byzantine nodes.
However, their scheme requires interactions between multiple nodes in a subtree to validate an aggregate contribution.

\subsection{Blockchain specific solutions}
Algorand~\cite{gilad2017algorand} is a proof-of-stake blockchain. Validation is delegated to a sizeable committee chosen from the validator set.
Contributions are circulated on a gossip network, and nodes on the gossip network must verify signatures before propagating them. The paper explicitly leaves the design of a Sybil-resistant gossip network for future work. Algorand's committee size is 2000, which they show to be safe for up to 20\% of Byzantine participants. This is already an important number of nodes, as today Bitcoin or Ethereum have 10000 nodes each, but necessary because the committee must contain at least 2 thirds + 1 honest nodes.
With these settings, Algorand measured a time to reach agreement (i.e. aggregating more than two thirds of votes) of 12 seconds. This should be contrasted with the sub-second average of Handel with 4000 nodes presented in section~\ref{sec:eval}.
Algorand secures its committees by means of \textit{cryptographic sortition}: committee members are secretly and randomly selected using a public source of randomness; committee members reveal themselves when sending their signatures. This secures them against DoS but does not secure the system against bribing. A one-to-one comparison is impossible as Algorand is a complete blockchain solution while Handel aims to be a building block with a limited but well defined scope.
Both solutions rely on shared public randomness. Algorand's cryptographic sortition allows for limited discrepancies in knowledge of membership.
Handel does not need a Sybil-resistant gossip network and remains efficient even in the presence of a large proportion of Byzantine nodes.

Avalanche~\cite{2019ScalableAP} is a proof-of-stake blockchain designed with the same objectives as Handel: execution times on the order of a second without using committees. It includes a binary consensus protocol called \textit{Snowflake}. Its design has also some similarities with Handel: Avalanche nodes communicate directly rather than through a peer-to-peer network protocol.
Like Handel, Avalanche uses randomness: at each round nodes randomly select a subset of the nodes they will be communicating with. An interesting difference is the communication mode. In Handel, nodes push their contributions to the other nodes. Avalanche uses a  \textit{query} mode, where nodes can request contributions from one another. As a consequence, Avalanche has no need for a windowing mechanism, as nodes can decide which nodes to request contributions from. A second consequence is a higher sensitivity to timeouts. When nodes don't respond to queries, Avalanche nodes wait for a timeout to expire before moving on. Lastly, Avalanche communication requires a round-trip, while in Handel communication is one-way.
\section{Conclusion}\label{sec:conclusion}
Handel improves on existing aggregation frameworks suitable for deployment with fail-silent nodes by doing away with timeouts and by supporting heterogeneous network and CPU capacities.
This versatility is a necessary feature for integration with public proof-of-stake (PoS) blockchains, which, by definition, run on globally distributed unpermissioned networks of machines of widely varying capabilities.
Handel's Byzantine-tolerance and speed
allows PoS blockchains to operate without committees, thus increasing their security,  and still achieve time to finality on the order of seconds.

\section*{Acknowledgements}
We thank Justin Cappos for his helpful insights,
Vanessa Bridge, Franck Cassez, Ben Edgington, Shahan Khatchadourian, Peter Robinson and Roberto Saltini for their comments and suggestions, Nedim Haveric for sharing his expertise on AWS, Dieter Mitsche and Nicolas Broutin for helpful discussions, and Christine Bégassat for careful proof-reading.

\nocite{*}

\bibliographystyle{ieeetr}
\bibliography{main}

\begin{thebibliography}{10}

\bibitem{gilad2017algorand}
Y.~Gilad, R.~Hemo, S.~Micali, G.~Vlachos, and N.~Zeldovich, ``{Algorand:
  scaling Byzantine agreements for cryptocurrencies},'' in {\em Proceedings of
  the 26th Symposium on Operating Systems Principles}, pp.~51--68, ACM, 2017.

\bibitem{hanke2018dfinity}
T.~Hanke, M.~Movahedi, and D.~Williams, ``Dfinity technology overview series,
  consensus system,'' {\em arXiv preprint arXiv:1805.04548}, 2018.

\bibitem{buchman2016tendermint}
E.~Buchman, ``{Tendermint: Byzantine fault Tolerance in the age of
  blockchains},'' 2016.

\bibitem{eth2}
{Ethereum Foundation}, ``{Ethereum 2}.''
  \url{https://github.com/ethereum/eth2.0-specs}, 2019.

\bibitem{canetti2001adaptive}
R.~Canetti, I.~Damgaard, S.~Dziembowski, Y.~Ishai, and T.~Malkin, ``On adaptive
  vs. non-adaptive security of multiparty protocols,'' in {\em International
  Conference on the Theory and Applications of Cryptographic Techniques},
  pp.~262--279, Springer, 2001.

\bibitem{Bonneau2016WhyBW}
J.~Bonneau, ``Why buy when you can rent? - bribery attacks on bitcoin-style
  consensus,'' in {\em Financial Cryptography Workshops}, 2016.

\bibitem{McCorry2018SmartCF}
P.~McCorry, A.~Hicks, and S.~Meiklejohn, ``Smart contracts for bribing
  miners,'' in {\em Financial Cryptography Workshops}, 2018.

\bibitem{Micali2001}
S.~Micali, K.~Ohta, and L.~Reyzin, ``Accountable-subgroup multisignatures,'' 01
  2001.

\bibitem{boldyreva2002}
A.~Boldyreva, ``Threshold signatures, multisignatures and blind signatures
  based on the gap-diffie-hellman-group signature scheme,'' in {\em PKC},
  pp.~31--46, 01 2003.

\bibitem{boneh2001short}
D.~Boneh, B.~Lynn, and H.~Shacham, ``{Short signatures from the Weil
  pairing},'' in {\em International Conference on the Theory and Application of
  Cryptology and Information Security}, pp.~514--532, Springer, 2001.

\bibitem{gupta2001scalable}
I.~Gupta, R.~Van~Renesse, and K.~P. Birman, ``Scalable fault-tolerant
  aggregation in large process groups,'' in {\em 2001 International Conference
  on Dependable Systems and Networks}, pp.~433--442, IEEE, 2001.

\bibitem{kempe2003gossip}
D.~Kempe, A.~Dobra, and J.~Gehrke, ``Gossip-based computation of aggregate
  information,'' in {\em 44th Annual IEEE Symposium on Foundations of Computer
  Science, 2003. Proceedings.}, pp.~482--491, IEEE, 2003.

\bibitem{yalagandula2004sdims}
P.~Yalagandula and M.~Dahlin, ``A scalable distributed information management
  system,'' in {\em SIGCOMM}, 2004.

\bibitem{li2005implementing}
J.~Li, K.~Sollins, and D.-Y. Lim, ``Implementing aggregation and broadcast over
  distributed hash tables,'' {\em ACM SIGCOMM Computer Communication Review},
  vol.~35, no.~1, pp.~81--92, 2005.

\bibitem{castro1999}
M.~Castro and B.~Liskov, ``Practical byzantine fault tolerance,'' in {\em
  OSDI}, 1999.

\bibitem{blanchard2017}
P.~Blanchard, E.~M.~E. Mhamdi, R.~Guerraoui, and J.~Stainer, ``Machine learning
  with adversaries: Byzantine tolerant gradient descent,'' in {\em NIPS}, 2017.

\bibitem{cappos2008san}
J.~Cappos and J.~H. Hartman, ``San ferm{\'i}n: Aggregating large data sets
  using a binomial swap forest,'' in {\em NSDI}, 2008.

\bibitem{carpen2015performance}
M.~Carpen-Amarie, P.~Marlier, P.~Felber, and G.~Thomas, ``A performance study
  of java garbage collectors on multicore architectures,'' in {\em Proceedings
  of the Sixth International Workshop on Programming Models and Applications
  for Multicores and Manycores}, pp.~20--29, ACM, 2015.

\bibitem{Dimiduk2012HBaseIA}
N.~Dimiduk, A.~Khurana, and M.~H. Ryan, ``{HBase} in action,'' 2012.

\bibitem{chang2008bigtable}
F.~Chang, J.~Dean, S.~Ghemawat, W.~C. Hsieh, D.~A. Wallach, M.~Burrows,
  T.~Chandra, A.~Fikes, and R.~E. Gruber, ``Bigtable: A distributed storage
  system for structured data,'' {\em ACM Transactions on Computer Systems
  (TOCS)}, vol.~26, no.~2, p.~4, 2008.

\bibitem{handellib}
B.~Kolad, N.~Gailly, and N.~Liochon, ``Handel implementation in go.''
  \url{https://github.com/ConsenSys/handel}, 2019.

\bibitem{bn256cf}
Cloudflare, ``{BN256 Cloudflare's implementation}.''
  \url{https://github.com/cloudflare/bn256}, 2018.

\bibitem{barreto2005pairing}
P.~S. Barreto and M.~Naehrig, ``Pairing-friendly elliptic curves of prime
  order,'' in {\em International Workshop on Selected Areas in Cryptography},
  pp.~319--331, Springer, 2005.

\bibitem{mikuli}
B.~Kolad, ``{Benchmarking of Milagro BLS implementation}.''
  \url{https://github.com/ConsenSys/mikuli}, 2018.

\bibitem{dwork1988consensus}
C.~Dwork, N.~Lynch, and L.~Stockmeyer, ``Consensus in the presence of partial
  synchrony,'' {\em Journal of the ACM (JACM)}, vol.~35, no.~2, pp.~288--323,
  1988.

\bibitem{MurtaTM06}
C.~D. Murta, P.~R.~T. Jr., and P.~Mohapatra, ``Characterizing quality of time
  and topology in a time synchronization network,'' in {\em Proceedings of the
  Global Telecommunications Conference, 2006. {GLOBECOM} '06, San Francisco,
  CA, USA, 27 November - 1 December 2006}, 2006.

\bibitem{latencyInternet}
W.~Network, ``Global ping statistics.'' \url{https://wondernetwork.com/pings},
  Dec 2018.

\bibitem{wittgenstein}
V.~Bridge, B.~Kolad, and N.~Liochon, ``Wittgenstein: protocol simulator.''
  \url{https://github.com/ConsenSys/wittgenstein}, 2018.

\bibitem{concentrationChernoff}
H.~Chernoff, ``A measure of asymptotic efficiency for tests of a hypothesis
  based on the sum of observations,'' {\em Ann. Math. Statist. 23 (1952), no.
  4, 493--507. doi:10.1214/aoms/1177729330.}, 1952.

\bibitem{concentration_lecture_notes}
M.~Goemans, ``Chernoff bounds, and some applications.''
  \url{https://math.mit.edu/~goemans/18310S15/chernoff-notes.pdf}.

\bibitem{dingledinetor}
R.~Dingledine, N.~Mathewson, and P.~F. Syverson, ``Tor: The second-generation
  onion router,'' in {\em USENIX Security Symposium}, 2004.

\bibitem{cifuentes2016poor}
F.~Cifuentes, A.~Hevia, F.~Montoto, T.~Barros, V.~Ramiro, and
  J.~Bustos-Jim{\'e}nez, ``Poor man's hardware security module (pmhsm): A
  threshold cryptographic backend for dnssec,'' in {\em Proceedings of the 9th
  Latin America Networking Conference}, pp.~59--64, ACM, 2016.

\bibitem{i2p}
I2P, ``{Invisible Internet Project}.'' \url{https://geti2p.net/en/}, 2000.

\bibitem{clarke1999distributed}
I.~Clarke, O.~Sandberg, B.~Wiley, and T.~W. Hong, ``Freenet: A distributed
  anonymous information storage and retrieval system,'' in {\em International
  Workshop on Designing Privacy Enhancing Technologies: Design Issues in
  Anonymity and Unobservability}, (Berlin, Heidelberg), pp.~46--66,
  Springer-Verlag, 2001.

\bibitem{torDoSDefense}
R.~Jansen, ``{New Tor denial of service attacks and defenses}.''
  \url{https://blog.torproject.org/new-tor-denial-service-attacks-and-defenses},
  2014.

\bibitem{torPerf}
T.~Metrics. \url{https://metrics.torproject.org/torperf.html}, 2018.

\bibitem{liu2005linkable}
J.~K. Liu and D.~S. Wong, ``Linkable ring signatures: Security models and new
  schemes,'' in {\em International Conference on Computational Science and Its
  Applications}, pp.~614--623, Springer, 2005.

\bibitem{fujisaki2008traceable}
E.~Fujisaki and K.~Suzuki, ``Traceable ring signature,'' {\em IEICE
  transactions on fundamentals of electronics, communications and computer
  sciences}, vol.~91, no.~1, pp.~83--93, 2008.

\bibitem{naehrig2010new}
M.~Naehrig, R.~Niederhagen, and P.~Schwabe, ``New software speed records for
  cryptographic pairings,'' in {\em International Conference on Cryptology and
  Information Security in Latin America}, pp.~109--123, Springer, 2010.

\bibitem{gobGolang}
Golang, ``Gob encoding.'' \url{https://golang.org/pkg/encoding/gob/}, 2014.

\bibitem{jelasity2005gossip}
M.~Jelasity, A.~Montresor, and O.~Babaoglu, ``Gossip-based aggregation in large
  dynamic networks,'' {\em ACM Transactions on Computer Systems (TOCS)},
  vol.~23, no.~3, pp.~219--252, 2005.

\bibitem{van2003astrolabe}
R.~Van~Renesse, K.~P. Birman, and W.~Vogels, ``Astrolabe: A robust and scalable
  technology for distributed system monitoring, management, and data mining,''
  {\em ACM transactions on computer systems (TOCS)}, vol.~21, no.~2,
  pp.~164--206, 2003.

\bibitem{conedht}
R.~Bhagwan, G.~Varghese, and G.~M. Voelker, ``{Cone: augmenting DHTs to support
  distributed resource discovery},'' tech. rep., University of California, San
  Diego, July 2003.

\bibitem{van2004willow}
R.~Van~Renesse and A.~Bozdog, ``{Willow: DHT, aggregation, and
  publish/subscribe in one protocol},'' in {\em International Workshop on
  Peer-to-Peer Systems}, pp.~173--183, Springer, 2004.

\bibitem{grumbachdht}
S.~Grumbach and R.~Riemann, ``{Secure and trustable distributed aggregation
  based on Kademlia},'' {\em CoRR}, vol.~abs/1709.03265, 2017.

\bibitem{maymounkov2002kademlia}
P.~Maymounkov and D.~Mazieres, ``Kademlia: A peer-to-peer information system
  based on the xor metric,'' in {\em International Workshop on Peer-to-Peer
  Systems}, pp.~53--65, Springer, 2002.

\bibitem{2019ScalableAP}
``Scalable and probabilistic leaderless {BFT} consensus through
  metastability,'' 2019.

\end{thebibliography}

\clearpage

\appendices


\section{Proof of Convergence}\label{app: proof of convergence}

\subsection{Notations}

For nonnegative integers $k,j$, set $I_{k,j}=\llbracket j2^k,(j+1)2^k\llbracket$.
Thus $\#I_{k,j}=2^k$ and $I_{k,2j}\sqcup I_{k,2j+1}=I_{k+1,j}$.
Recall from section~\ref{sec:handel:tree} the notation $\Part{i}{k}$ for $i$'s level $k$ peer set. Observe that $\Part{i}{k}=I_{k,m}$ for some $m$.
We write $\calP(X)$ for the powerset of a finite set $X$, and $\calP_W(X)$ for the subset of cardinality $W$ subsets of $X$. For $q\in[0,1]$ and $m$ positive integer, we let $\calB(m,q)$ denote the binomial distribution with $m$ independent trials and success probability $q$. We use the shorthand $\calB(q)=\calB(1,q)$.

\subsection{Mathematical model and main result}
To simplify the analysis, we assume the number of participants $N$ is a power of two: $N=2^n$.
Every participant carries a unique ID in $\llbracket0,N\llbracket$
and we identify nodes with their ID.
In line with section~\ref{sec:handel:state}, we assume a public and unbiasable source of randomness to initially shuffle node IDs.
We thus consider a family of independent identically distributed (\textit{iid}) Bernoulli trials $(B_i)_{0\leq i<N}$, $B_i\sim\calB(b)$ for some $b\in[0,1]$, that determine if a node is Byzantine or not: $i$ is Byzantine if and only if $B_i=1$.
We let $\mathrm{HN}=\lbrace i\mid B_i=0\}$ be the set of honest nodes, and $\mathrm{HN}_{k,j}=\mathrm{HN}\cap I_{k,j}$.

We assume that every node $i$ independently and uniformly draws publicly computable levelwise Verification Priorities (\vpl$_i$), see section~\ref{sec:protocol:ranking}. From these we deduce, for every node $i$ and every level $k$, an $O(n)$-sized\footnote{$\outcomm$ for a well-chosen constant $C$} initial segment $\mathbf{L}^{i}_k\subset \Part{i}{k}$ of the \vpl$_i$, as well as an $O(n)$-sized\footnote{$2\outcomm$} initial segment of the Contribution Prioritization Vector.
\begin{thm_bis}[Handel Convergence]
\emph{Let $0<\byz_\mathrm{max}<1$ be an upper bound on the expected proportion of Byzantine nodes and let $0<\tau<1-\byz_\mathrm{max}$ be a target threshold.
There is a positive $C=C(\byz_\mathrm{max}, \tau)$ such that if honest nodes only read messages sent by the $\outcomm$ first nodes in every levelwise \vpl{}, and send at most $2\outcomm$ messages per level, then
\[
	\proba\left[
	\begin{array}{c}
	\text{all honest nodes aggre-}\\
	\text{gate a complete signature}
	\end{array}
	\right]
	\xrightarrow[\text{exp. fast}]{~n\to+\infty~}1
\]%
}
\end{thm_bis}

\subsection{Proof strategy}

The proof distinguishes two phases of communication: homogenization and reproduction. The \textbf{homogenization phase} sees nodes communicate with \textit{all} their low level peers, i.e. up to some level $\ell\ll n$. To achieve $O(n)$ message complexity, we take $\ell=\ln_2(n)+O(1)$. The purpose of homogenization is to achieve, with overwhelming probability, homogeneous peersets $I_{\ell,j}$, $0\leq j<2^{n-\ell}$. By this we mean that the proportion of Byzantine nodes in any of the $I_{\ell,j}$ is $\leq(1+\delta)\byz_\mathrm{max}$ and in particular $\frac{\#\mathrm{HL}_{k,\ell}}{2^\ell}\geq\tau$. Here $\delta>0$ is chosen so that $\tau\leq1-(1+\delta)\byz_\mathrm{max}$. When this is the case, all honest nodes assemble a valid level $\ell$-aggregate signature containing $\geq \tau 2^\ell$ contributions.

The \textbf{reproduction phase} concerns the levels $>\ell$: honest nodes communicate with those nodes $j$ whose levelwise \vpl$_j$ ranks them high, say among the first $\outcomm$ peers, and only verify incoming contributions from nodes in the chosen initial segment of the appropriate \vpl{}. To ensure success during the reproduction phase, one must make sure that (1) the \vpl{} of every honest node at every level $>\ell$ contains an honest node (2) no node is contained in ``too many'' \vpl{}s, say more than $2\outcomm$. If these conditions are met, every honest node will be contacted at every level $>\ell$ by an honest node in its \vpl{}, and won't need to send more than $2\outcomm=O(n)$ messages.

If both homogenization and reproduction phases succeed, then all honest nodes manage to assemble a threshold meeting signature (actually, a complete signature). We are therefore interested in bounding the probability that either homogenization or reproduction fail. We rely on the following \textbf{concentration inequality} for the binomial distribution \cite{concentrationChernoff, concentration_lecture_notes}
where $\mu=mq$ is the mean:
\begin{equation}\label{eq:concentration inequality excess}
    \forall\delta>0,~\proba\big[\calB(m,q)\geq (1+\delta)\mu\big]\leq\left(\frac{e^\delta}{(1+\delta)^{1+\delta}}\right)^{\mu}.
\end{equation}
We will freely use the fact that for events $A_1,\dots,A_m$,
$\proba(\cup_{i=1}^mA_i)\leq\sum_{i=1}^m\proba(A_i)$.

\subsection{Homogenization}

Let $(B_i)_{0\leq i<N}$ be \textit{iid} Bernoulli trials $\sim\calB(b)$ where $0\leq b\leq b_{\mathrm{max}}$.
Let $0\leq \ell\leq n$ be integer and let $0\leq j<2^{n-\ell}$.
We say that a batch $I_{\ell,j}$ is \textbf{homogeneous} if it contains no more than $(1+\delta)\cdot b_{\mathrm{max}}2^\ell$ Byzantine nodes; we define \textbf{success for homogenization} as the event that all batches  $I_{\ell,j}$, $0\leq j<2^{n-\ell}$ are homogeneous.
By \eqref{eq:concentration inequality excess}, the probability that $I_{\ell,j}$ is \textit{inhomogeneous} satisfies
\begin{IEEEeqnarray*}{RRL}
\IEEEeqnarraymulticol{3}{l}{
\proba\Big[
I_{\ell,j}\text{ inhomogeneous}
\Big]
=
\proba\bigg[
\sum_{i\in I_{\ell,j}}
B_i>(1+\delta)\cdot b_{\mathrm{max}}2^\ell
\bigg]}\\
& \qquad= &
\proba\big[
\calB(2^\ell, b)
>(1+\delta)\cdot b_{\mathrm{max}}2^\ell
\big]
\\
& \leq &
\proba\big[
\calB(2^\ell, b_{\mathrm{max}})
>(1+\delta)\cdot b_{\mathrm{max}}2^\ell
\big]
\\
& \leq &
\left(
1-\theta
\right)^{b_{\mathrm{max}}2^\ell}
\end{IEEEeqnarray*}
where $1-\theta = \frac{e^\delta}{(1+\delta)^{1+\delta}}\in\,]0,1[$. If $r$ is a nonnegative integer such that $\ln(2) - b_{\mathrm{max}}|\ln(1-\theta)|2^r<-1~({\color{ForestGreen}\bm{\star}})$, then by setting $\ell = \lceil \ln_2(n)\rceil + r$, we have
\begin{IEEEeqnarray*}{RRL}
\IEEEeqnarraymulticol{3}{l}{
\proba\Big[
\text{Homogenization fails}
\Big]}\\
& = &
\proba\left[
\bigcup_{0\leq j<2^{n-\ell}}
\big[
I_{\ell,j}\text{ inhomogeneous}
\big]
\right]\\
& \qquad\leq &
\sum_{0\leq j<2^{n-\ell}}
\proba
\big[
I_{\ell,j}\text{ inhomogeneous}
\big]
\\
& = &
\sum_{0\leq j<2^{n-\ell}}\proba\bigg[
\sum_{i\in I_{\ell,j}}
B_i>(1+\delta)\cdot b_{\mathrm{max}}2^\ell
\bigg]\\
& \leq &
2^{n-\ell}(1-\theta)^{b_{\mathrm{max}}2^\ell}
\\
& \leq &
\exp\Big(\ln(2)n - b_{\mathrm{max}}|\ln(1-\theta)|2^\ell\Big)
\leq 
\exp\big(-n\big)
\end{IEEEeqnarray*}
Thus homogenization succeeds with probability $\geq 1-e^{-n}$ which tends exponentially fast to $1$.

\subsection{Reproduction}

Let $C>0$ be a positive constant to be determined later. We define \textbf{reproductive success} for a node $i$ at level $k\geq \ell$ as the event that (1) its level $k$ \vpl{} contains an honest node and (2) the number of level $k$ nodes whose \vpl{} contains $i$ is $\leq 2\outcomm$. We consider \textbf{success of reproduction} the event that all honest nodes, at all levels $\geq\ell$ have had reproductive success.

\subsubsection{Likelihood of a VPL containing no honest node.}
Let $\mathbf{W}\in\calP(I_{k,j})$ be subset with $\frac{\#\mathbf{W}}{2^k}\geq\tau_0>0$. Let $\mathbf{L}$ be a uniformly sampled cardinality $\outcomm$-subset of $I_{k,j}$. Then
\begin{IEEEeqnarray*}{RRL}
\IEEEeqnarraymulticol{3}{l}{\proba
\Big[
\mathbf{L}
\,\cap\,
\mathbf{W}
=
\emptyset
\Big]
=
\frac{\binom{2^k-W}{\outcomm}}{\binom{2^k}{\outcomm}}}\\
& \qquad= &
\frac{2^k-\outcomm}{2^k}\cdot
\frac{2^k-1-\outcomm}{2^k-1}
\cdots
\frac{2^k-W+1-\outcomm}{2^k-W+1}\\
& \leq &
\Big(1-\frac{\outcomm}{2^k}\Big)^{W}
\leq
\exp\big(-\tau_0\outcomm\big)
\leq
\exp\big(-\tau_0 Cn\big)
\end{IEEEeqnarray*}

\subsubsection{Likelihood of a node being in $\geq2\outcomm$ many \vpl{}s}

Let $0\leq i<N$ and $m$ be such that $i\in I_{k,m}$.
Then $I_{k,m}\sqcup \Part{i}{k}=I_{k+1,\lfloor m/2\rfloor}$.
Let $(\mathbf{L}_k^j)_{j\in \Part{i}{k}}$ be \textit{iid} random variables with uniform distribution in $\calP_{\outcomm}(\Part{i}{k})$. Let $j\in \Part{i}{k}$: the probability that $\mathbf{L}_k^j$ contains $i$ is
\[
    \proba
	\Big[
	i\in\mathbf{L}_k^j
	\Big]
	=
	\frac{\binom{2^k-1}{\outcomm-1}}{\binom{2^k}{\outcomm}} = \frac{\outcomm}{2^k}
\]
The random variable $\#\{j\in I_{k,m}\mid i\in\mathbf{L}_k^j\}$ thus follows the binomial distribution $\calB(2^k,\frac{\outcomm}{2^k})$. Therefore, applying \eqref{eq:concentration inequality excess} with $\delta=1$,
\begin{IEEEeqnarray*}{RRL}
\IEEEeqnarraymulticol{3}{l}{%
\proba
\Big[
\#\big\{j\in I_{k,m}\text{ s.t. }i\in\mathbf{L}_k^j\big\}
\geq 2\outcomm
\Big]}\\
& \qquad= & 
\proba
\Big[
\calB\Big(2^k,\frac{\outcomm}{2^k}\Big)
\geq 2\outcomm
\Big]\leq
\left(\frac{e}{4}\right)^{\outcomm}
\leq\left(\frac{e}{4}\right)^{Cn}
\end{IEEEeqnarray*}

\subsubsection{Probability that reproduction fails}

Let us set $\tau_0=1-(1+\delta)b_\mathrm{max}\geq\tau$.
Suppose that homogenization is a success. Then for all 
$k\geq \ell$ and all $0\leq j<2^{n-k}$, $\#\big(\mathrm{HN}\cap I_{k,j}\big)\geq \tau_0 2^k$, and so
\begin{IEEEeqnarray*}{RRL}
\IEEEeqnarraymulticol{3}{l}{
\proba\Big[
\text{Reproduction fails}
\Big]
=
\proba\left[
\bigcup_{\substack{\ell\leq k<n\\i\in\mathrm{HN}}}
\left[
\begin{array}{c}
    i\text{ reproductive}  \\
    \text{failure at level } k
\end{array}
\right]
\right]}\\
& \qquad\leq &
\sum_{\substack{\ell\leq k<n\\ i\in\mathrm{HN}}}
\proba
\left[
\begin{array}{l}
    \Big[
    \mathbf{L}^i_k\cap\big(\Part{i}{k}\cap\mathrm{HN}\big)=\emptyset
    \Big]
    \\
    \quad\cup\;
    \Big[
        \#\big\{j\in \Part{i}{k}\text{ s.t. }i\in\mathbf{L}^j_k\big\}
        \geq 2\outcomm
    \Big]   
\end{array}
\right]
\\
& \leq &
n2^n\Big(\exp\big(-\tau_0 Cn\big) + \left(\frac{e}{4}\right)^{Cn}\Big)
\end{IEEEeqnarray*}

\subsection{Probability that homogenization and reproduction succeed}

Consider the event $\mathcal{S}$ where both homogenization and reproduction succeed. We can now show that for the appropriate choice of $C$, $\proba[\mathcal{S}^c]$ tends exponentially fast to $0$ as $n\to+\infty$. Indeed,
\begin{IEEEeqnarray*}{RCL}
\proba\Big[
\mathcal{S}^c
\Big]
& = &
\proba\left[
\big[
\text{Homogenization fails}
\big]
\cup
\big[
\text{Reproduction fails}
\big]
\right]\\
& \leq &
\exp\big(-n\big)
+
n2^n\Big(
\exp\Big(-\tau_0 Cn\Big)
+
\left(\frac{e}4\right)^{ Cn}
\Big)
\end{IEEEeqnarray*}
Now letting $C>\max\Big\{\frac{\ln(2)}{\tau_0}, \frac{\ln(2)}{2\ln(2)-1}\Big\}~({\color{ForestGreen}\bm{\star\star}})$\footnote{Since $\tau_0\geq\tau>0$, $C>\max\Big\{\frac{\ln(2)}{\tau}, \frac{\ln(2)}{2\ln(2)-1}\Big\}$ is enough}, the right hand side tends to $0$ exponentially fast.
\begin{rmk}
The above proof adapts to obtain the following (sharper) result: for arbitrary $\alpha>0$, the failure probability $\proba\big[\mathcal{S}^c\big]$ can be made $\leq N^{-\alpha}$ asymptotically while retaining the same polylogarithmic complexity by appropriately picking larger values for the constants $r$ and $C$ in equations $({\color{ForestGreen}\bm{\star}})$ and $({\color{ForestGreen}\bm{\star\star}})$.
\end{rmk}


\end{document}